\documentclass[notitlepage,10pt,aps,prd,twocolumn,amsmath,amssymb,floatfix,superscriptaddress,nofootinbib,preprintnumbers]{revtex4-2}

\usepackage{amsfonts,amssymb,amsmath,graphicx,color,bm,epsfig}
\usepackage[normalem]{ulem}
\usepackage{multirow}
\usepackage{centernot}

\definecolor{ultramarine}{rgb}{0.07, 0.04, 0.56}
\definecolor{cadmiumgreen}{rgb}{0.0, 0.42, 0.24}
\definecolor{indigo(dye)}{rgb}{0.0, 0.25, 0.42}
\usepackage[linktocpage=true]{hyperref}
\hypersetup{
colorlinks=true,
citecolor=ultramarine,
linkcolor=cadmiumgreen,
urlcolor=indigo(dye),
}

\def\GeV{\ {\rm GeV}}
\def\MeV{\ {\rm MeV}}
\def\TeV{\ {\rm TeV}}

\newcommand{\Mpl}{M_{\rm Pl}}

\def\higgs{{h}}
\def\hard{{\rm hard}}
\def\soft{{\rm soft}}

\def\QCD{{\rm QCD}}

\def\mphi{m_\phi}
\def\Gphi{\Gamma_\phi}
\def\TRH{T_{\rm RH}}

\def\Hend{H_{\rm end}}
\def\HFO{H_{\rm FO}}
\def\Hunstable{H_{\rm unstable}}
\def\hend{h_{\rm end}}

\def\HdS{H_{\rm dS}}

\def\psplit{p_{\rm split}}
\def\psplithiggs{p_{\rm split}^{\rm higgs}}

\def\pin{p_{\rm in}}
\def\pout{p_{\rm out}}
\def\pLPM{p_{\rm LPM}}

\def\Gtilde{\tilde{\Gamma}}

\def\Gsplit{\Gamma_{\rm split}}
\def\Gel{\Gamma_{\rm el}}

\def\Nh{\mathcal{N}}
\def\Nhdec{\Nh_{h\text{-decay}}}
\def\Ndec{N_{h\text{-decay}}}

\def\Tphi{T_\phi}
\def\Tmax{T_{\phi\text{-max}}}
\def\Tmaxhiggs{T_{h\text{-max}}}
\def\Ntherm{N_{\phi\text{-therm}}}
\def\Nthermhiggs{N_{h\text{-therm}}}

\def\rhoparam{\tilde{\rho}_h}

\def\hconf{\underline{h}}

\def\xFO{x_{\rm FO}}

\DeclareRobustCommand{\rchi}{{\mathpalette\irchi\relax}}
\newcommand{\irchi}[2]{\raisebox{\depth}{$#1\chi$}} 
\def\chiconf{\underline{\rchi}}
\def\osc{\textrm{osc}}

\newcommand\numberthis{\addtocounter{equation}{1}\tag{\theequation}}
\definecolor{darkgreen}{cmyk}{0.85,0.2,1.00,0.2} 
\definecolor{purple}{cmyk}{0.5,1.0,0,0} 

\begin{document}

\title[]{The Secret Higgstory of the Highest Temperature during Reheating}

\author{Samuel Passaglia}
\email{samuel.passaglia@ipmu.jp}
\affiliation{Kavli Institute for the Physics and Mathematics of the Universe (WPI),The University of Tokyo Institutes for Advanced Study (UTIAS),The University of Tokyo, Chiba 277-8583, Japan}

\author{Wayne Hu}
\affiliation{Kavli Institute for Cosmological Physics, Department of Astronomy \& Astrophysics, 
Enrico Fermi Institute, University of Chicago, Chicago, IL 60637, U.S.A.}

\author{Andrew J. Long}
\affiliation{Department of Physics and Astronomy, Rice University, Houston, Texas 77005, U.S.A.}

\author{David Zegeye}
\affiliation{Kavli Institute for Cosmological Physics, Department of Astronomy \& Astrophysics, 
Enrico Fermi Institute, University of Chicago, Chicago, IL 60637, U.S.A.}

\label{firstpage}

\begin{abstract}

We study the role of the Standard Model Higgs condensate, formed during cosmological inflation, in the epoch of reheating that follows. We focus on the scenario where the inflaton decays slowly and perturbatively, so that there is a long period between the end of inflation and the beginning of radiation domination. The Higgs condensate decays non-perturbatively during this period, and we show that it heats the primordial plasma to much higher temperatures than would result from the slowly-decaying inflaton alone. We discuss the effect of this hot plasma on the thermalization of the inflaton's decay products, and study its phenomenological implications for the formation of cosmological relics like dark matter, with associated isocurvature fluctuations, and the restoration of the electroweak and Peccei-Quinn symmetries.

\end{abstract}

\date{\today}
\maketitle 

\section{Introduction}
\label{sec:intro}

The discovery of the Higgs boson in 2012 not only provided the final piece of the Standard Model of particle physics, but also stimulated the realization that the Higgs is of fundamental significance to cosmology. Its deep connections to cosmic inflation, and in particular its dynamics during inflation, have attracted special attention \cite{EliasMiro:2011aa,Kearney:2015vba,Espinosa:2015qea,East:2016anr,Espinosa:2017sgp,Lu:2019tjj,Passaglia:2019ueo}.

If the Higgs is a light spectator field during inflation \cite{Chernikov:1968zm}, then its quantum fluctuations accumulate on superhorizon scales and  locally displace the field away from the minimum of its potential \cite{Linde:1982uu,Affleck:1984fy,Lee:1987qc,Starobinsky:1994bd}. A Higgs condensate is formed, 
which does not survive in the universe today but rather is destroyed during the epoch of reheating, when the Higgs and inflaton transferred their condensate energy to a thermal bath of Standard Model 
particles.  

In this paper, we present a new physical prediction for how reheating of the Standard Model proceeds when the inflaton field $\phi$ decays {\it slowly}, meaning that its decay rate $\Gphi$ is small enough that
\begin{equation}
    \label{eq:Gtilde}
    \Gtilde \equiv \frac{\Gphi}{\mphi^3} \Mpl^2 \ll 1 \;.
\end{equation}
Here $\mphi$ is the inflaton's mass, and $\Mpl =(8\pi G)^{-1/2}$
%
is the reduced Planck mass. If the inflaton decays, for example, through a dimension-5 Planck-suppressed operator $\mathcal{L} = (\phi/ \Mpl) \mathcal{O}_4 $, then the decay rate is parametrically $\Gphi \sim \mphi^3 / \Mpl^2$ and $\Gtilde \sim 1$. Models with $\Gtilde \ll 1$ then occur when the coupling that mediates the decay of the inflaton to matter is weaker than gravity \cite{Ellis:2015pla}.

A tiny coupling is technically natural, and couplings of gravitational strength or weaker for the inflaton are in general desirable on theoretical grounds to avoid spoiling the flatness of the inflaton's potential \cite{Assassi:2013gxa,Ellis:2015pla}, and on phenomenological grounds to avoid overproducing gravitinos during reheating \cite{Kawasaki:2008qe}. 

In this regime, the decay of the inflaton is a perturbative process which can be understood analytically. As the inflaton decays to relativistic particles during reheating, it  sources the primordial plasma with an energy density $\rho_r$ evolving according to 
\begin{equation}
    \frac{d\rho_r}{d N} + 4 \rho_r = 3 \Gtilde \, \mphi^3  H
\end{equation}
until reheating is completed. 
We use the $e$-folds $N\equiv\ln a$ counted from the end of inflation as our time coordinate throughout this work, and $H = H(N)$ is the Hubble parameter.

Under the assumption that the inflaton's potential is quadratic near its minimum, the inflaton condensate's coherent oscillations drive an effectively matter-dominated cosmological expansion with $H \propto e^{-3N/2}$.  
The radiated energy density then has the asymptotic solution 
\begin{equation}
    \label{eq:rho_decay_products}
    \rho_r = \frac{6}{5} \, \Gtilde \, \mphi^3 \, H \;.
\end{equation}
As the Hubble rate drops, this radiation loses energy in absolute terms but gains energy relative to the inflaton condensate itself. 

Reheating is  generally taken to complete when $\Gamma_\phi \sim H$, at a time
\begin{equation}
    N_{\rm RH} \sim \frac{2}{3} \ln\left( \frac{\Hend}{\mphi^3} \frac{\Mpl^2}{\Gtilde} \right)\;,
\end{equation}
where $\Hend$ denotes the Hubble rate at the end of inflation. The plasma temperature at this time, i.e.\ the reheating temperature, is taken by convention as \cite{Chung:1998rq,Giudice:2000ex}
\begin{equation}
    \label{eq:TRH}
    \TRH  = \left( \frac{90}{\pi^2 g_*} \right)^{1/4} \sqrt{\frac{\Gtilde \mphi^3}{\Mpl} } \;,
\end{equation}
where $g_*$ is the effective number of relativistic degrees of freedom of the plasma. 

Since the radiation energy density decreases throughout the reheating epoch, the maximum temperature of the thermal component of the plasma can be much larger than $\TRH$ \cite{Chung:1998rq,Giudice:2000ex}. If the Higgs condensate is neglected, the maximum temperature during reheating is achieved when a majority of the radiation produced by the inflaton has thermalized, giving the relation \cite{Harigaya:2013vwa}
\begin{align}
    \label{eq:Tmaxinflaton}
    \Tmax \simeq \left.\left( \frac{30 \rho_r}{\pi^2 g_*} \right)\right\vert_{\Ntherm}^{1/4} \;, 
\end{align}
where the thermalization time $\Ntherm$ can be computed as a function of $\Gtilde$ \cite{Mukaida:2015ria}.  
Since the radiation energy density evolves as $\rho_r \propto H \propto e^{-3N/2}$ during matter domination, the maximum plasma temperature scales as $\Tmax \propto e^{-3\Ntherm/8}$.
We mark $\Ntherm$ and $\Tmax$ with $\phi$'s to denote they are associated with the decay of the inflaton. 

In this work, we show that $\Tmax$ is not necessarily the maximum temperature of the primordial plasma during reheating. The decay of the Higgs condensate formed during inflation can lead to a higher temperature.

After inflation the Higgs condensate oscillates around the minimum of its potential. We assume that the amplitude of the Higgs condensate is not so large as to probe the classically unstable part of its potential. Avoiding this instability requires either that the energy scale of inflation is below $\Hunstable \sim 10^{10} \GeV$ or that the Higgs potential is stabilized at high energies \cite{Espinosa:2015qea,Kohri:2016wof,Adshead:2020ijf}.
As the Higgs condensate oscillates in its approximately quartic potential, its energy density redshifts like radiation.  Therefore, well after the end of inflation, the Higgs' energy density can be simply parameterized as
\begin{equation}
    \label{eq:rhohiggs}
    \frac{\rho_\higgs}{\Hend^4} = \rhoparam e^{-4 N} \;,
\end{equation}
where $\rhoparam$ is a constant, dimensionless parameter which encodes all the details of how the Higgs condensate was formed and how it began to oscillate.  
We will show that it is generally of order unity or larger. The Hubble rate at the end of inflation, $\Hend$, will mainly scale all of our results together rather than affect the relative importance of the various processes.

After a few oscillations, the Higgs condensate decays by parametric resonance in a completely Standard Model process \cite{Enqvist:2015sua,Figueroa:2015rqa}.  This produces an effectively thermal plasmas at some time $\Nthermhiggs$ not long after the end of inflation, with temperature
\begin{align}
    \label{eq:Tmaxhiggs}
    \Tmaxhiggs  \simeq \left.\left( \frac{30 \rho_\higgs}{\pi^2 g_*} \right)\right\vert_{\Nthermhiggs}^{1/4} 
    \;.
\end{align}

Comparing the maximum temperature of the Higgs condensate's contribution to the plasma \eqref{eq:Tmaxhiggs} to the maximum temperature of the inflaton condensate's contribution \eqref{eq:Tmaxinflaton}, we see that even though the radiation from the inflaton will eventually dominate the energy density of our universe, the radiation from the Higgs controls the maximum temperature of our universe if
\begin{equation}
    \label{eq:Tmaxcomparison}
    \frac{\Tmaxhiggs}{\Tmax} 
    \sim \left( \frac{\rhoparam}{\Gtilde} \left(\frac{\Hend}{\mphi}\right)^3 \right)^{1/4} e^{ 3\Ntherm/8 -\Nthermhiggs}
\end{equation}
exceeds unity. We denote parametric relations with $\sim$ here and throughout.
With $\rhoparam$ generically of order unity or larger, and $\mphi\sim\Hend$ as is typical for relatively large field inflation\footnote{One would generally expect that if inflation ended on a quadratic potential  $\Hend/m_\phi \sim \phi_{\rm end}/\Mpl$.}
we see that the Higgs contribution dominates over the inflaton condensate contribution so long as it sources a thermal population sufficiently before the inflaton does and $\Gtilde$ is small. Decreasing $\Gtilde$ both suppresses the energy density of the inflaton decay products and delays their thermalization time $\Ntherm$. Technical naturalness allows it to be very small, and empirically it is constrained only by ensuring that reheating completes before big bang nucleosynthesis 
around $1 \MeV$  \cite{Kawasaki:2000en,deSalas:2015glj,Hasegawa:2019jsa}, yielding a lower bound,
\begin{equation}
    \label{eq:Gtildemin}
    \Gtilde \gtrsim 10^{-17} \left( \frac{\TRH }{1 \MeV} \right)^{2} \left( \frac{10^{10} \GeV}{\mphi } \right)^{3}  \left( \frac{g_* }{100 } \right)^{1/2}\;.
\end{equation}

We will therefore be able to show that for much of the allowed parameter space, the maximum temperature of our universe is provided by the decay of the Higgs condensate after reheating, with our results joining a growing body of work highlighting the importance of the Standard Model Higgs in a wide variety of reheating scenarios:  when the inflaton decays quickly \cite{Freese:2017ace,Litsa:2020rsm}; when the inflaton decays to a hidden sector \cite{Tenkanen:2019cik} or has a stiff post-inflationary equation of state \cite{Figueroa:2016dsc}; and when the condensate relaxation contributes to leptogenesis \cite{Kusenko:2014lra,Yang:2015ida}.

This article is organized as follows. Table~\ref{tab:symbols} follows this introduction and summarizes the main parameters of this work and gives the equations in which they are defined.

In \S\ref{sec:higgs}, we track the Higgs condensate from its formation during inflation, through its oscillations during reheating, to its resonant decay to effectively thermal Standard Model radiation. We estimate the parameters $\rhoparam$ and $\Nthermhiggs$ which control the maximum temperature of the Higgs contribution to the primordial plasma.

In \S\ref{sec:inflaton}, we review how the inflaton condensate decays to a population of underoccupied hard particles, slowing down thermalization, to relate the inflaton-decay product thermalization time $\Ntherm$ to the inflaton decay rate $\Gtilde$. We compute for the first time the effect of the Higgs decay products on the thermalization of the inflation decay products. 

In \S\ref{sec:comparison} we compare the Higgs and inflaton contributions to the plasma and show that for much of the parameter space the maximum temperature of our universe is controlled by the Higgs. We show that this temperature will generally have inhomogeneities on large scales uncorrelated with the adiabatic fluctuations sourced by the inflaton.

In \S\ref{sec:implications}, we discuss the theoretical and observational implications of having a maximum temperature controlled by the Higgs, which include the enhancement of the production of relic particles and the restoration of spontaneously broken symmetries.
We conclude in \S\ref{sec:conclusion}.

\begin{table}[t!]
\begin{center}
\begin{tabular}{ |c|l|c| }
\hline
Symbol & Gloss & Eqn. \\
\hline \hline

\hline
& {\textbf{Inflaton control parameters}} &  \\
$\Hend$ & Hubble rate at the end of inflation & \eqref{eq:rhohiggs} \\
$\Gtilde$ & Inflaton decay rate & \eqref{eq:Gtilde} \eqref{eq:Gtildemin}\\
$\mphi$ & Inflaton mass 
& \eqref{eq:Gtilde} \eqref{eq:Tmaxcomparison} \\

\hline
& {\textbf{Higgs control parameters}} &  \\
$\rhoparam$ & Dimensionless asymptotic Higgs energy & \eqref{eq:rhohiggs} \eqref{eq:typical_rhoparam} \\
$\Nthermhiggs$ & Higgs effective thermalization time & \eqref{eq:Tmaxhiggs} \eqref{eq:typical_Ndec} \\

\hline
& {\textbf{Auxiliary derived parameters}} &  \\
$\TRH$ & Reheating temperature & \eqref{eq:TRH} \\
$\Ntherm$ & Inflaton decay products therm. time & \eqref{eq:Tmaxinflaton} \eqref{eq:Ntherm} \\

\hline
&{\textbf{Maximum temperatures}} &  \\
$\Tmax$ & Maximum $T$ from inflaton decay 
& \eqref{eq:Tmaxinflaton} 
\eqref{eq:Tphimax_over_Hend}\\
$\Tmaxhiggs $ & Maximum $T$ from Higgs decay & \eqref{eq:Tmaxhiggs} \eqref{eq:typicalT} \\

\hline

\end{tabular}
\caption{The parameters and outputs of the slow reheating scenario in the Standard Model.}
\end{center}
\label{tab:symbols}
\end{table}

\section{Radiation from the Higgs}
\label{sec:higgs}

In this section, we discuss the formation of the Higgs condensate during inflation, its dynamics at the end of inflation, and its decay to effectively thermal radiation via parametric resonance after inflation. 

\subsection{Condensate formation}
\label{subsec:stochastic}
In their pioneering work, the authors of Ref.~\cite{Starobinsky:1994bd} studied the equilibrium state of a self-interacting scalar field in a de~Sitter background.  
This is done by treating the field amplitude, coarse-grained on a fixed physical scale larger than the Hubble scale $\sim 1/H$, as a random variable.  
Its evolution is governed by a competition between deterministic rolling and stochastic fluctuations of amplitude $\sim H/2\pi$ per $e$-fold as the exponentially-expanding vacuum fluctuations cross the averaging scale \cite{Starobinsky:1986fx}. 

The probability distribution over field amplitudes can be calculated by finding the stationary solutions of a Fokker-Planck equation.  For example in the case of four real scalar fields with an $\mathrm{SO}(4)$ symmetry in their quartic self-interaction $\lambda \vec{\varphi}^4/4$, the equilibrium distribution in de~Sitter space with Hubble parameter $\HdS$ has moments $\langle \vec{\varphi} \rangle = 0$ and~\cite{Adshead:2020ijf} 
\begin{equation}
    \label{eq:equilibrium_rms_adshead}
    \sqrt{\langle h^2 \rangle}
    = \left( \frac{3}{8 \pi} \right)^{1/4} \frac{\HdS}{\lambda^{1/4}}
    \;,
\end{equation}
where $h \equiv |\vec{\varphi}|$. We say then that $h$ forms a scalar condensate. 

This discussion carries over to the $\mathrm{SU}(2)$-doublet Higgs field during inflation.  
Assuming that the inflationary Hubble scale is much larger than the electroweak scale $v \simeq 246 \ \mathrm{GeV}$, the Higgs field will develop a scalar condensate.  
The typical condensate amplitude can be estimated in de~Sitter using Eq.~\eqref{eq:equilibrium_rms_adshead} upon identifying $\lambda$ with the Higgs self coupling \cite{Kunimitsu:2012xx,Espinosa:2015qea}. 
For simplicity, we neglect any running of $\lambda$ and in numerical estimates take $\lambda = 0.01$. We assume throughout that the Higgs is minimally coupled to gravity and not directly coupled to the inflaton.  

The cosmological inflationary epoch is only quasi-de~Sitter, and the Hubble parameter decreases as inflation proceeds. Near the end of inflation in particular, the Hubble rate can evolve quickly enough that a condensate established stochastically early during inflation can further evolve and the de~Sitter result~\eqref{eq:equilibrium_rms_adshead} becomes an inaccurate estimate of the typical condensate amplitude \cite{Hardwick:2017fjo,Fumagalli:2019ohr}.
This evolution can be tracked directly by solving the Fokker-Planck equation, but it can also be estimated by noting that it is dominated by deterministic rolling rather than stochastic fluctuations. 

The local behavior of the condensate amplitude away from the origin, where we can ignore the effective angular momentum barrier associated with non-radial fluctuations, can then be understood from the Klein-Gordon equation of motion 
\begin{equation}
\label{eq:eom_nonoise}
\frac{d^2 h}{d N^2} + \frac{3}{2} (1-w) \frac{d h}{dN} + \frac{V_{,h}}{H^2}  = 0\;,
\end{equation}
where $w$ is the background equation of state; $w\simeq -1$ during inflation and $w=-1/3$ when inflation ends. 

When the Hubble drag conditions $|dw/dN|\ll1$ and $|d \ln V_{,h}/dN| \ll1$ are satisfied, the Klein-Gordon equation admits slow-roll solutions satisfying \cite{Gordon:2004ez}
\begin{equation}
    \label{eq:drag}
    \frac{d h}{d N} \simeq \frac{-2}{3 (3+w)} \frac{V_{,h}}{H^2}
    \;,
\end{equation}
but the drag conditions break down if the condensate amplitude is larger than
\begin{equation}
    \label{eq:saturation}
    h \sim \frac{H}{\sqrt{\lambda}}
    \;,
\end{equation}
which corresponds to the point where the Higgs field's effective mass $\sim \sqrt{\lambda} h$ is comparable to the Hubble scale $H$.

Therefore when the Hubble rate drops significantly near the end of inflation, the condensate can be released from Hubble drag if $h \gtrsim H / \sqrt{\lambda}$. The condensate then rolls down the potential until it is again halted by Hubble drag when $h/\Hend \sim \lambda^{-1/2}$, and the Higgs displacement at the end of inflation is then independent of the Hubble rate earlier in inflation.

When this occurs the  Higgs at the end of inflation will have a smaller displacement than the equilibrium distribution in the earlier phase of inflation would suggest, but a larger displacement than the equilibrium distribution with $\HdS=\Hend$ would imply. This is consistent with the results of Refs.~\cite{Hardwick:2017fjo,Fumagalli:2019ohr} which solved the Fokker-Planck equation for a spectator field in a variety of inflationary backgrounds.

\begin{figure}[t]
\includegraphics[]{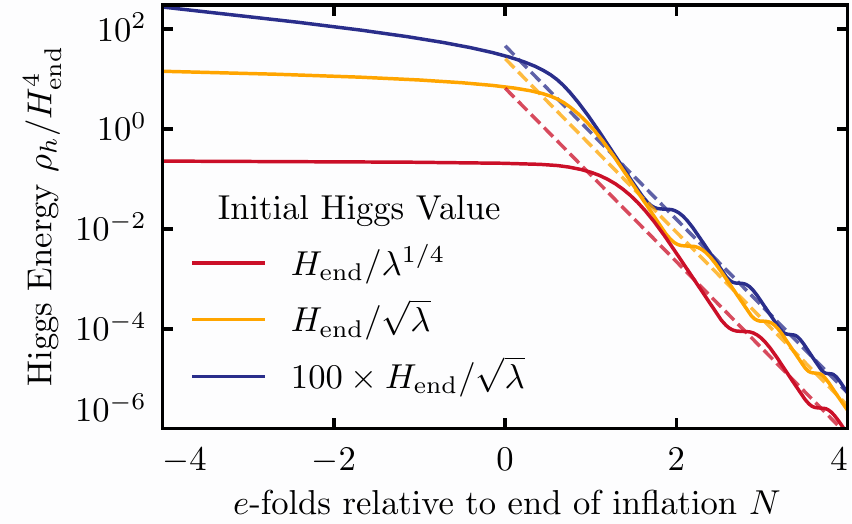}
\caption{The evolution of Higgs fluctuations through the end of $m_\phi^2 \phi^2$ inflation. No matter how large fluctuations are early in inflation (purple), they roll down as the Hubble rate decreases and end up bounded by the drag value \eqref{eq:saturation} at the end of inflation (yellow). This is larger than the equilibrium value for fluctuations in de~Sitter with Hubble set by $\Hend$ (red). The asymptotic energy density after inflation (dashed) depends only on $\hend^{4/3}$ rather than $\hend^4$ because small fluctuations remain frozen for a longer period after inflation.}
\label{fig:energy_density_during}
\end{figure}

We show this effect in Fig.~\ref{fig:energy_density_during} by solving the equation of motion \eqref{eq:eom_nonoise} with various initial conditions near the end of $\mphi^2 \phi^2$ inflation. 
In this inflationary model the Hubble rate at the end of inflation has decreased by an order of magnitude from the Hubble rate when CMB scales crossed the horizon, i.e.\ $\Hend/H_{-60} \simeq 1/12$ and $\mphi/\Hend \simeq 2$.  Due to the dynamics discussed above, the condensate's amplitude at the end of inflation is roughly $\hend \sim \Hend/\sqrt{\lambda}$, even if the initial displacement was much larger.

Therefore though the Higgs field amplitude at the end of inflation 
$\hend/\Hend$ 
is a stochastic variable, it should be within an order unity factor of the typical values
\begin{align}
    \label{eq:range_for_hend}
    \frac{\hend}{\Hend} \sim  
    \min\left[ \frac{\HdS}{\Hend} \lambda^{-1/4}, \lambda^{-1/2}\right]\;,
\end{align}
where the first argument corresponds to the case where the Hubble rate does not decrease significantly near the end of inflation, and the second argument is the case where it does. $\HdS$ is now identified with the Hubble rate when the Higgs departed from the equilibrium solution. Since $\HdS/\Hend>1$ in any inflationary model, this window of typical expectations is fairly narrow in practice, at most a factor of $\sim\lambda^{-1/4} \simeq 3$ for $\lambda \sim 10^{-2}$.
 
\begin{figure*}[t]
\includegraphics[]{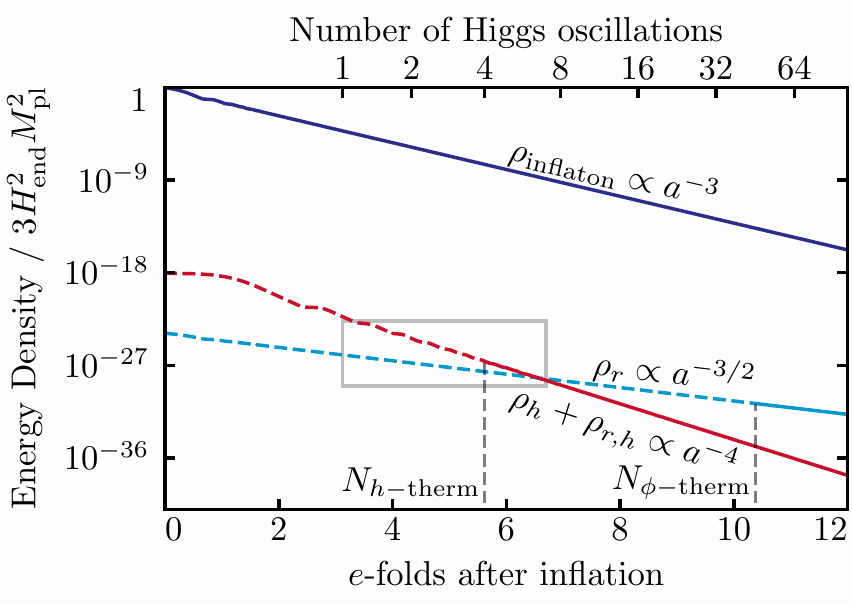}
\includegraphics[]{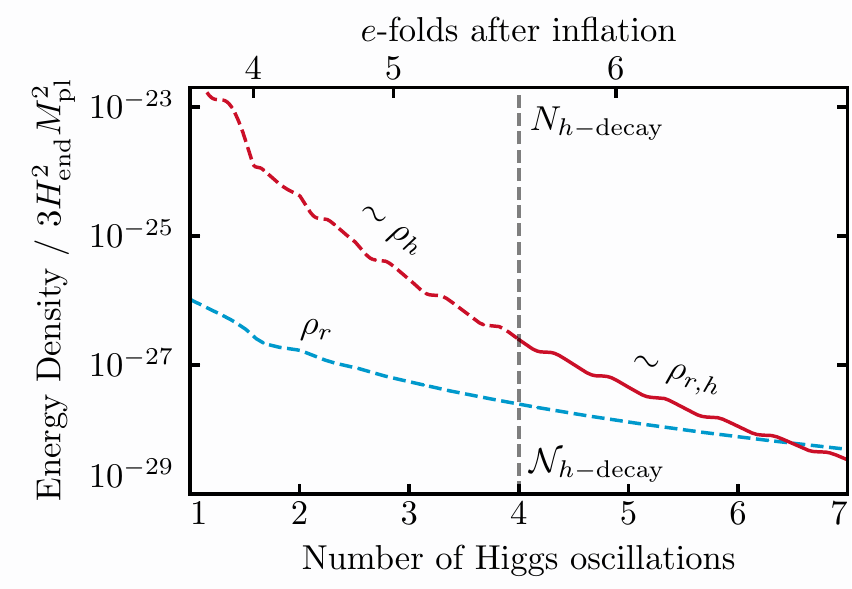}
\caption{After inflation, the Higgs condensate's energy density (red, $\rhoparam = 10$) is negligible compared to the inflaton's (purple), but it can be larger than the energy of the inflaton's decay products (blue, $\Gtilde= 10^{-7}$). The Higgs condensate (dashed) thermalizes (solid) some $\Nthermhiggs\sim6$ $e$-foldings after inflation, which can be well before the inflaton condensate's decay products thermalize at $\Ntherm$. The Higgs condensate thermalization time is associated with the condensate's decay at $\Ndec$ (boxed region, shown in right panel), which takes some $\Nhdec\sim 4$ Higgs amplitude oscillations. We have assumed here that inflation ends in a $\mphi^2 \phi^2$ potential 
with $\mphi=10^{10} \GeV$, and we have not explicitly modeled the decay of the Higgs condensate here so $\rho_{r, h}$ continues to oscillate after $\Ndec$.}
\label{fig:energy_density}
\end{figure*}

After the end of inflation at $N=0$, the Hubble rate drops and the condensate will eventually be released from Hubble drag and oscillate in its quartic potential. As seen in Fig.~\ref{fig:energy_density_during}, in this regime the condensate energy density redshifts on the cycle average like radiation \cite{Greene:1997fu}, 
\begin{equation}\label{eq:rhohOfN}
\begin{split}
    \frac{\rho_\higgs}{\Hend^4} 
    = \rhoparam e^{-4N} \;,
\end{split}
\end{equation}
which defines $\rhoparam$ as the dimensionless, time-independent asymptotic energy of the condensate.
We calibrate $\rhoparam$ directly as a function of $\hend/\Hend$ 
by solving the equation of motion \eqref{eq:eom_nonoise} numerically starting from the end of inflation with Hubble-dragged initial velocity and assuming that the inflaton potential is quadratic near the end of inflation and thereafter, $V = \mphi^2 \phi^2 / 2$. 
From solutions in the range $\hend/\Hend \in [0, \lambda^{-1/2}]$, we infer an empirical fitting function for the asymptotic Higgs energy 
\begin{equation}
\begin{split}
\label{eq:rho_param}
\rhoparam
& \sim \frac{1}{4} \lambda \left(\frac{ \hend}{\Hend}\right)^4 \times \left(\frac{\Hend}{\sqrt{\lambda}\hend}\right)^{8/3} \times 1.3
\; \\
& \simeq 0.33 \, \lambda^{-1/3} \left(\frac{ \hend}{\Hend}\right)^{4/3}\;.
\end{split}
\end{equation}
The first factor of the first line accounts for the energy density carried by the Higgs field at the end of inflation, the second factor accounts for the Hubble drag phase until $H \sim \sqrt{\lambda} \hend$ during which the energy density is approximately constant rather than redshifting like $e^{-4N}$, and the final numerical factor 1.3 is calibrated from the numeric solutions. This fitting function agrees parametrically with the fitting functions provided for a wide range of post-inflationary backgrounds by Ref.~\cite{Figueroa:2015rqa}. 

For the typical Higgs displacements at the end of inflation given in \eqref{eq:range_for_hend}, we therefore have
\begin{equation}
    \label{eq:typical_rhoparam} 
    \rhoparam \sim
    \min\left[
   {\lambda^{-2/3}}\left( \frac{\HdS}{\Hend}\right)^{4/3}, {\lambda^{-1}}\right] \;.
\end{equation}
In our examples and to normalize scaling relations we take $\rhoparam = 10$ reflecting $\HdS \sim \Hend$ and $\lambda \sim 10^{-2}$. Similarly the asymptotic Hubble rate after inflation ends and the inflaton oscillates on a quadratic potential can be approximated as
\begin{equation}
    \label{eq:H}
    \frac{H}{\Hend} \simeq 0.8 \, e^{-3N/2}
    \;,
\end{equation}
where the coefficient is a fit to numerical results.

The solution with $\rhoparam=10$ is shown in the first several $e$-folds after the end of inflation in Fig.~\ref{fig:energy_density}. 
We compare this Higgs condensate energy density to the energy density of the inflaton and its decay products \eqref{eq:rho_decay_products} for an inflaton decay rate with $\Gtilde = 10^{-7}$. While the Higgs condensate's energy density is much smaller than that of the inflaton, it can be large in comparison to the inflaton's decay products. 
\subsection{Condensate decay}

After inflation, we assume that the Higgs condensate is on a purely radial trajectory, which we have checked is a good approximation due to Hubble friction. Rather than $|\vec{\varphi}|$, $h$ now represents a field that oscillates around zero and its equation of motion asymptotically approaches the simple form \ \cite{Greene:1997fu}
\begin{equation}\label{eq:Higgs_eqn_of_motion}
    \ddot{\hconf} + \lambda \hconf^3 = 0
    \;,
\end{equation}
where $\hconf \equiv e^N h$ is the conformally rescaled field and overdots denote derivatives with respect to the conformal time  $\eta = \int dt/e^{N}$. This asymptotic equation of motion has a solution in terms of an elliptic cosine function
~\cite{Greene:1997fu} 
\begin{equation}\label{eq:h_analytic}
    \hconf \simeq \hconf_\osc \,  \textrm{cn}\bigl(\sqrt{\lambda} \hconf_\osc (\eta-\eta_\osc), \tfrac{1}{2}\bigr)
    \;,
\end{equation}
which is periodic in $z(\eta) 
\equiv \sqrt{\lambda} \hconf_\osc \eta,$ with period $\Delta z \equiv \Gamma(1/4)^2/\sqrt{\pi} \simeq 7.4$ where $\Gamma(x)$ is the Euler gamma function.  $\hconf_\osc$ is the amplitude of the conformally conserved oscillations, and $\eta_\osc$ is an arbitrary turning point $\dot{\hconf}(\eta_\osc) = 0$.
Matching this solution to the asymptotic energy density \eqref{eq:rhohiggs} relates $\hconf_\osc$ to $\rhoparam$ as 
\begin{equation}
\label{eq:hconfosc}
\frac{\hconf_\osc}{\Hend} =\left(\frac{4 \rhoparam}{\lambda}\right)^{1/4} \;.
\end{equation}

The Higgs condensate oscillates with this amplitude until it decays into Standard Model particles~\cite{Enqvist:2013kaa,Enqvist:2013qba,Figueroa:2014aya,Figueroa:2015rqa,Enqvist:2015sua} via parametric resonance~\cite{Kofman:1994rk,Kofman:1997yn,Greene:1997fu}. 
The most efficient decay channel is to weak gauge bosons \cite{Figueroa:2015rqa}, 
since they have a large coupling to the Higgs field and they do not experience Pauli blocking~\cite{GarciaBellido:2000dc}. 
Following Refs.~\cite{Enqvist:2013qba,Figueroa:2015rqa}, the Higgs decay to one such field $A 
\in \left\{ Z, W^\pm \right\}$ can be modeled using a set of $3$ scalar fields $\chi_i$, one for each helicity, and a corresponding scalar-Higgs interaction $\mathcal{L}_\mathrm{int} = - g_A^2 \vec{\chi}^2 h^2 / 8$ which is familiar from studies of inflationary preheating~\cite{Kofman:1994rk,Kofman:1997yn,Greene:1997fu,Amin:2014eta}. This modeling provides an qualitative understanding of the Higgs decay by parametric resonance analytically.

The equation of motion for the mode functions $\rchi_k$ in the oscillatory regime is~\cite{Greene:1997fu} 
\begin{equation}
\label{eq:modefunction_eom}
\chiconf_k'' + \left(\left(\frac{k}{ \sqrt{\lambda} \hconf_\osc}\right)^2 + q \left(\frac{\hconf}{\hconf_\osc}\right)^2 \right) \chiconf_k = 0
\;,
\end{equation}
where the conformal mode function $\chiconf_k \equiv e^N \rchi_k$, the resonance parameter $q\equiv g_A^2 / (4 \lambda)$,  and $'$ denotes a derivative with respect to $z$.  
For a given energy scale, the Higgs self coupling $\lambda$ and the various gauge couplings can be computed in the Standard Model, with typical values $\lambda \sim 10^{-2}$ and $g_Z^2\sim 0.6$, $g_W^2\sim0.3$ yielding $q\sim 10$. 

At early times the Higgs condensate's energy loss from decay can be neglected, $\hconf$ follows the analytic solution  \eqref{eq:h_analytic}, and the mode equation \eqref{eq:modefunction_eom} describes a Lam\'{e} equation which exhibits resonant behavior and exponentially-growing solutions for certain ranges of $k / (\sqrt{\lambda} \hconf_\osc)$ and $q$.  
When $n(n+1)/2 < q < (n+1)(n+2)/2$ for an odd integer $n$,  the first resonance band extends from $0$ to $k_\ast \sim (\sqrt{\lambda} \hconf_\osc) (q / (2 \pi^2))^{1/4}$ ~\cite{Greene:1997fu}.  

Modes in the resonance band grow exponentially in $z$ and yield an exponentially increasing occupation number of particles
\begin{equation}
\label{eq:fA_def}
    f(k,z) = \frac{1}{2} \left(e^{2 \mu_{k} (z-z_\osc)}-1\right) \;,
\end{equation}
with $z_\osc \equiv z(\eta_\osc)$ and with the Floquet exponent $\mu_{k}$ a non-monotonic function of $k$ and $q$ bounded by $2 \ln\left(1+\sqrt{2}\right) / \Delta z \simeq 0.24$. 
It can be approximated as a top hat
\begin{equation}
    \label{eq:floquet}
    \mu_{k} \sim \vert\mu\vert \Theta(1-k/k_* )\;,
\end{equation}
with $\vert\mu \vert \sim 0.2$ \cite{Figueroa:2015rqa}. The boson $A$ is now identified with the weak gauge boson which yields the largest Floquet exponent, since it will dominate the condensate decay.
Its exponentially increasing number density then yields an energy density 
\begin{align}
 \rho_A &\simeq \frac{\hconf_\osc^4 q^{5/4} \lambda^2}{2^{13/4} \pi^{7/2}} e^{-4(N-N_\osc)} \left(e^{2 |\mu| (z - z_\osc)}-1\right) \;.
\end{align}
where in the per-particle energy we have accounted for the effective mass  $g^2 \langle h^2 \rangle / 4 \simeq g^2 h_{\rm osc}/8 $ from the condensate displacement.
A key timescale is $z_{h\text{-decay}}$ when the condensate has transferred an $\mathcal{O}(1)$ fraction of its energy to the gauge boson, $\rho_A = \rho_\higgs$.
This time is conveniently expressed in terms of a number of Higgs field oscillations, 
\begin{equation}
\begin{split}
\label{eq:Nhdecay}
    \Nhdec 
    & \equiv (z_{h\text{-decay}}-z_\osc) / \Delta z \\  
    & = \frac{1}{2 |\mu|} \frac{1}{\Delta z} \ln\left[1+\frac{2^{5/4} \pi^{7/2}}{q^{5/4} \lambda}\right]\;.
\end{split}
\end{equation}
Using the fiducial values provided above, we estimate that the Higgs condensate decays after $\Nhdec \sim 2$ field oscillations. This number depends only logarithmically on all of the parameters in the problem except the Floquet exponent $|\mu|$, and is independent of the energy in the Higgs condensate.\footnote{One might wonder whether a resonance analysis is valid for such a small number of oscillations --- just $4$ zero crossings --- but numerical solutions validate it within an intrinsic uncertainty of $1$ zero crossing \cite{Greene:1997fu,Enqvist:2014tta}.}

This simple analytic estimate that the Higgs decays in just a few oscillations is well supported by the lattice simulations performed by Refs.~\cite{Figueroa:2015rqa,Enqvist:2015sua}, which incorporate respectively abelian and non-abelian gauge structures for the Higgs. They find that the condensate decays most of its energy after $\Nhdec = 3\sim4$ oscillations.

To absorb the theoretical uncertainty in this number, we treat the number of Higgs field oscillations until decay $\Nhdec$ as free parameter in our reheating study, with the expectation that it lies in the above range.

Keeping $\Nhdec$ a free parameter also allows us to absorb the scenario where none of the Standard Model weak gauge bosons exhibit a $k\rightarrow 0$ resonance. In this case, while there will always be a resonance band at some $k$ and in the large $q$ limit the resonance parameter $\mu_k$ in that band will still approach the maximum possible value $\simeq 0.24$, in general the Floquet exponent $\mu_k$ in a higher resonance band will be smaller and imply a correspondingly longer decay time $\Nhdec$.  Refs.~\cite{Figueroa:2015rqa, Enqvist:2015sua} find that backreaction can take up to $\sim 4$ times more oscillations than in the usual case.

Finally, we need the number of cosmological $e$-folds that elapse between the end of inflation and the time of Higgs condensate decay.
In the oscillatory regime the number of $e$-folds $\Delta N$ in an interval of $\Delta \Nh$ oscillations is
\begin{equation}
    \label{eq:Nh}
    \Delta N \simeq 2 \ln\left(\Delta \Nh \Delta z/2 + 1\right) \;.
\end{equation}
To get the absolute number of $e$-folds  $N$ in terms of the absolute number of oscillations $\Nh$, we calibrate the mapping empirically by numerical solution of the condensate equation of motion. Identifying $\Nh$ as half the number of zero-crossings of the field, the number of $e$-folds to reach $\Nhdec$ for a condensate exiting inflation with dimensionless displacement $\hend/\Hend$ is 
\begin{equation}
\begin{split}
\label{eq:Ndec}
    \Ndec & \sim  2 \ln\left(\frac{ \Nhdec \Delta z}{2}\right) 
    \\ & \quad 
    -\frac{2}{3} \ln\left(\frac{\sqrt{\lambda}\hend}{\Hend}\right) 
- 0.4 \,,
\end{split}
\end{equation}
where  the first term accounts for the $e$-folds from the beginning of oscillations until decay 
\eqref{eq:Nh}, the second for the $e$-folds between the end of inflation and the beginning of the Higgs oscillations, and the final numerical term is calibrated from the numeric solutions. For $\Nhdec=4$, the typical decay time is
\begin{equation}
\label{eq:typical_Ndec}
\Ndec \sim \ln\left( \frac{140}{\sqrt{\lambda \rhoparam}} \right) \sim 6\;,
\end{equation}
where we have evaluated with $\rhoparam \sim 10$ and $\lambda \sim 0.01$.


\subsection{Effective Temperature and Thermalization}

We have seen that in just a few oscillations of the Higgs field, most of the Higgs condensate's energy is transferred to weak gauge bosons. This radiation is very abundant in the sense that the mode occupancy per helicity is much larger than one. Specifically, combining Eqs.~\eqref{eq:fA_def}~and~\eqref{eq:Nhdecay} yields an estimate of the gauge boson spectrum when the condensate decays.  The occupation number is
\begin{equation}\label{eq:weak_boson_spectrum}
     f_* \equiv f(p, \Nhdec) \sim \frac{2^{1/4} \pi^{7/2}}{q^{5/4} \lambda
    } = \mathcal{O}(100)
\end{equation}
in a resonance band that extends up to a physical momentum  
\begin{equation}
    p_*
    \sim \left(\frac{q}{2 \pi^2}\right)^{1/4} \sqrt{\lambda} \hconf_\osc \, e^{-\Ndec} \;.
\end{equation}
The number density 
$n_* \sim f_* p_*^3$ and energy density 
$\rho_* \sim f_* p_*^4$ of produced particles is dominated by momenta $p \sim p_*$.  
This momentum is larger than the Hubble rate, 
\begin{equation}
\label{eq:hubble_decay}
    H (\Ndec) \sim \frac{p_*}{\Nhdec q^{1/4}}\;, 
\end{equation}
which confirms that a particle description is valid. 

The weak boson radiation is already near thermal at the time of production. If the 
energy of the decay products were redistributed into a thermal spectrum with $g_\ast$ relativistic species, then the typical energy per particle would be set by the temperature 
$T \sim (\rho_{h,*}/g_*)^{1/4}$. 
With $\langle E \rangle$ denoting the population average energy, comparing this expression to the per-particle energy of the produced particles yields
\begin{equation}
    \frac{\langle E \rangle_{h\text{-decay}}}{\langle E \rangle_{T}} \sim \frac{p_*}{f_*^{1/4} p_* g_*^{-1/4}} \sim \frac{g_*^{1/4}}{f_*^{1/4}} \sim 1 \;,
\end{equation}
and reveals that most of the energy is carried by particles with momenta $p_* \sim T$ comparable to the eventual temperature of the thermalized system.

With particles of similar number density and energy to those of a thermal distribution, the weak boson spectrum that results from parametric resonance is therefore effectively thermal. This observation motivates us to identify the effective thermalization time $\Nthermhiggs$ with the Higgs decay time $\Ndec$, i.e. 
\begin{equation}
\label{eq:Nthermhiggs}
\Nthermhiggs \simeq \Ndec\;.
\end{equation}
At this time the effective temperature of the Higgs' decay products follows by energy conservation,
\begin{equation}
\label{eq:typicalT}
\begin{split}
    \frac{\Tmaxhiggs}{\Hend} 
    & =  \left( \frac{30  \rhoparam}{\pi^2 g_*} \right)^{1/4} e^{- \Nthermhiggs} \\ 
    & \sim 10^{-3}  \left(\frac{\rhoparam}{10}\right)^{1/4}
    e^{-(\Nthermhiggs-6)}
    \;,
\end{split}
\end{equation}
where we have used $g_\ast \sim 100$ and scaled the result to typical values for $\rhoparam$ and $\Nthermhiggs$ from Eqs.~\eqref{eq:typical_rhoparam},~\eqref{eq:typical_Ndec},~and~\eqref{eq:Nthermhiggs}. 

As we will see in \S\ref{sec:inflaton}, this is in sharp contrast with the situation for the inflaton condensate, which decays by producing less abundant particles but with much larger energy than that of a thermal distribution of the same energy density. 

The subsequent evolution and complete thermalization of the effectively thermal plasma of Higgs decay products then depends on processes adjusting the number of particles and their momentum distribution. An important process while the Higgs is decaying is the non-linear interaction of the resonantly produced particles \cite{Lozanov:2019jxc,Lozanov:2017hjm}. The non-abelian interactions of the Higgs' decay products are especially efficient at extending the overoccupied particles at $p_*$ to higher momenta with order unity occupancy, which is even closer to thermal \cite{Enqvist:2015sua}. At low momentum, scattering and absorption/emission processes likewise bring the distribution closer to thermal \cite{Kurkela:2014tea}.

\section{Radiation from the Inflaton}
\label{sec:inflaton}

We now discuss the decay of the inflaton and the thermalization of its decay products. In contrast to the Higgs condensate's rapid decay via parametric resonance, which produces an abundance of effectively thermal particles, the inflaton's decay is perturbative, and its decay products first take the form of a severely underoccupied distribution of hard primaries. These slowly transfer their energy to a thermal soft population via in-medium splitting, delaying thermalization of the full energy released by the inflaton and lowering the maximum temperature of the inflaton's decay products. 

Our calculations are based on the extensive literature on thermalization of non-abelian plasmas \cite{Baier:2000sb, Arnold:2001ba,Arnold:2002zm}, applied to the cosmological context of reheating \cite{Davidson:2000er,Kurkela:2011ti,Harigaya:2013vwa,Kurkela:2014tea,Mukaida:2015ria}. We provide the first calculation of how this thermalization process proceeds in the presence of the decay products of the Higgs condensate.

\subsection{Inflaton condensate and decay}
After inflation, the energy density of the universe is dominated by the coherent oscillations of the inflaton condensate. Radiation domination begins once the condensate fully decays. Well before that time, the condensate's energy loss is negligible, and it can be treated as a free field oscillating in a quadratic potential $V(\phi) = \mphi^2 \phi^2 / 2$ with a mass parameter $\mphi$ corresponding to the mass of an inflaton particle. On the cycle average the inflaton's energy density $\rho_\phi \propto e^{-3N}$ corresponds to a matter-dominated universe $H \propto e^{-3N/2}$ as in Eq.~\eqref{eq:H}. Each inflaton particle in the cold condensate carries an energy of $\sim m_\phi$ and their number density is 
\begin{equation}
    n_\phi \simeq \rho_\phi / \mphi \simeq 3 \Mpl^2 H^2 / \mphi
    \;.
\end{equation}

If the inflaton decays at a rate $\Gphi = m_\phi^3 \Gtilde / \Mpl^2$ into 
relativistic particles, then the energy density $\rho_r$ of the emitted radiation obeys
\begin{equation}
    \frac{d\rho_r}{d N} + 4 \rho_r = \frac{\Gphi}{H} \rho_\phi \simeq 3 \Gtilde \mphi^3 H \;. 
\end{equation}
The asymptotic solution is
\begin{equation}
    \label{eq:rho_decay_products_bis}
    \rho_r \simeq \frac{6}{5} \, \Gtilde \, \mphi^3 \, H \;.
\end{equation}

\subsection{Hard primaries}


We assume that the inflaton decays to pairs of relativistic Standard Model particles, each with energy $\sim\mphi / 2$. 
The energy of each hard primary produced by the inflaton redshifts as $e^{-N}$.  
The number density of hard primaries obeys 
\begin{equation}
    \frac{dn_\hard}{dN}+3 n_\hard=2\Gphi\frac{n_\phi}{H} \;,
\end{equation}
where the factor of $2$ accounts for the pair of hard primaries produced by each inflaton particle decay.  
After some time their accumulated phase space distribution function is then~\cite{McDonald:1999hd,Allahverdi:2000ss} (also \cite{Harigaya:2013vwa,Mukaida:2015ria})  
\begin{equation}\label{eq:f_hard}
\begin{split}
    & f_\hard(p, N) = \frac{3 \cdot 2^{7/2} \pi^2}{g_{\hard}} \Gtilde \left( \frac{p}{\mphi} \right)^{-3/2} \frac{H}{\mphi} \\ 
    & \quad \text{for} \quad m_\phi e^{-N}/2 \lesssim p \lesssim \mphi / 2 
    \;,
\end{split}
\end{equation}
where $g_\mathrm{hard}$ counts the decay products' redundant internal degrees of freedom (e.g., color and spin). 
The numerical coefficients ensure that the energy density of hard particles is just $\rho_\hard = \rho_r$ from Eq.~\eqref{eq:rho_decay_products_bis}, while the number density evaluates to 
\begin{equation}\label{eq:rhoH_nH}
\begin{split}
    n_\hard & = 4\Gtilde \, \mphi^2 \, H
    \;.
\end{split}
\end{equation}
In these calculations, the momentum integration
is dominated by the UV cutoff $p = \mphi/2$, and we drop terms suppressed by $H/\Hend$ that are negligible after the end of inflation.  

Though the hard particles are sourced continuously, the energy and number density of the hard particle population is always dominated by those produced in the most recent Hubble time. The per-particle energy $\sim \mphi$ is then much larger than it would be in a thermal bath of the same energy $T \sim (\rho_r/g_*)^{1/4}$; explicitly,
\begin{equation}
\frac{\langle E \rangle_\hard}{\langle E \rangle_{T}} \sim \frac{\mphi}{(g_*^{-1} \Gtilde \, \mphi^3 \, H)^{1/4}} \gg 1 \;.
\end{equation}
This quantity is large because the total decay product energy is suppressed by $\Gtilde\ll1$ and decreases with time, while the energy of each particle is
$\sim \mphi$ which we assume is comparable to or larger than $\Hend$. Equivalently, the number density of particles in the hard population is much smaller than those in a thermal distribution of the same energy density.

The hard primaries are therefore very far from thermal. In order to thermalize, they must transfer their energy to an abundant population of soft particles.

\subsection{Energy cascade toward thermal bath}
\label{subsec:cascade}

The hard primaries emit lower momentum soft particles by collinear splitting in a medium comprised of the hard primaries themselves and the products of the previous splittings.  This splitting leads to a cascade of energy from an underoccupied hard distribution to an abundant soft population. The total energy density of the radiation (hard + soft) obeys Eq.~\eqref{eq:rho_decay_products_bis} until backreaction on the condensate occurs and reheating completes. For pedagogy and to connect to the existing literature, we neglect here 
 the influence of the decay products of the Higgs.  Their impact can change the thermal history of the plasma, as we discuss in \S\ref{subsec:higgsoninf}.

The abundantly populated soft particles rapidly thermalize at a temperature $T_\soft$ that is much less than the hard particle energy $\mphi/2$. Most of the energy $\rho_\soft$ and particle number $n_\soft$ in the soft population is carried by particles with momentum $p \sim T_\soft$. $T_\soft$ itself is determined by the energy the hard particles have lost through in-medium splitting. This transfer from hard to soft is a bottleneck that prevents immediate thermalization of the full energy released as the inflaton decays.

To write down the in-medium splitting rate, we must specify the nature of the hard particles and their interactions.  
For concreteness we consider a non-abelian plasma~\cite{Baier:2000sb}, and suppose that the hard primaries are gluons. Then $g_\hard = 16$ 
and the strong coupling is then denoted by $\alpha = g_s^2 / 4\pi \sim 0.1$.

By emitting soft radiation, a gluon of momentum $\pin$ can split and form a gluon of momentum $\pout \lesssim \pin$. The rate at which this splitting occurs in the medium is estimated as \cite{Bethe:1934za,Kurkela:2011ti}
\begin{equation} \label{eq:Gamma_split}
    \begin{split}
    & \Gsplit(N,\pin,\pout) \sim \alpha \Gel \min\left[1, \sqrt{\frac{\pLPM}{\pout}}  \right] \\ 
    & \qquad \text{for} \quad \pout \lesssim \pin 
    \;,
    \end{split}
\end{equation}
where $\Gel$ is the rate for elastic gluon scattering. $\Gsplit$ only depends on the incident gluon's momentum $\pin$ through the kinematic restriction $\pout \lesssim \pin$, and $\Gsplit$ is suppressed for high-momentum daughters $\pout > \pLPM$ by the Landau-Pomeranchuk-Migdal (LPM) effect~\cite{Landau:1953um,Migdal:1956tc}.
The LPM effect is the destructive interference of radiation produced from nearby scattering sites when the formation time for the radiation is long 
compared to the typical time between scatterings. In QCD this occurs for daughter particles with momentum $\pout$ above \cite{Baier:1996kr,Zakharov:1996fv} (see also Refs.~\cite{Arnold:2001ba,Arnold:2001ms,Arnold:2002ja})
\begin{equation}
    \pLPM \sim \frac{m^2}{\Gel}\;,
\end{equation}
where $m$ is the gluon's screening mass in the medium.  We estimate 
\begin{equation}
\begin{split}
    m^2 &\sim \alpha \sum_{\QCD}{\int \frac{f}{p}  d^3 p} \\ &\sim \alpha \frac{n_\hard}{\mphi} + \alpha \frac{g_\QCD}{g_*} \frac{n_\soft}{T_\soft} \;,
\end{split}
\end{equation}
which has contributions from the hard particles and from the thermal soft population, containing $g_\QCD \sim 88$ quark and gluon degrees of freedom.
The screening mass also acts as an infrared regulator for the elastic scattering rate
\begin{align*}
    \label{eq:Gamma_elastic}
    \Gel &\sim \int \! \! d^2 q \frac{\alpha^2}{q^2 (q^2 + m^2)} \sum_\QCD{\int \! \! d^3 p f(p) (1\pm f(p))}\\
    &\sim \frac{\alpha^2}{m^2}\left(n_\hard + \frac{g_\QCD}{g_*} n_\soft \right)\;. \numberthis
\end{align*}

At this point we remind the reader that all expressions containing the coupling $\alpha$ should be viewed parametrically.  For example we neglect numerical factors like the quadratic Casimir, and we neglect logarithmic factors that appear when solving the exact system of Boltzmann equations describing collinear splitting in a non-abelian plasma. 

Energy transfer between the hard primaries and the soft population relies on efficiently producing low-momentum particles by splitting.  
This splitting is efficient if $\Gsplit \gtrsim H$.
Particles with large momenta $\pout$ are more difficult to produce due to the LPM suppression, and their production is inefficient above a momentum scale $\psplit$ where
\begin{equation}
    \Gsplit(\psplit(N), N) \equiv H(N)\;,
\end{equation}
which implies 
\begin{equation}\label{eq:p_split}
\begin{split}
    \psplit & = \psplit^\hard + \psplit^\soft \\ 
    & \sim \alpha^4 \left( \frac{n_\hard}{H^2} + \frac{g_\QCD}{g_*} \frac{n_\soft}{H^2} \right) 
    \;.
\end{split}
\end{equation}
Note that $\psplit$ receives contributions from collisions with both the hard and soft particles, and in \S\ref{subsec:higgsoninf} we will include the contribution from collisions with the decay products of the Higgs.

Since the splitting rate \eqref{eq:Gamma_split} does not significantly depend on the incident momentum $\pin$, any secondaries produced with $\pout < \psplit$ can themselves radiate efficiently. When they do so they lose an order unity fraction of their energy, and in this way the energy cascades from the hard particles to the thermal bath.

\subsection{Asymptotic temperature of the bath}
\label{subsec:soft}
The energy cascading down from the hard particles accumulates in the thermal bath of soft particles, such that the energy density of the bath obeys
\begin{equation}
    \frac{d\rho_\soft}{d N} + 4 \rho_\soft =  \int_0^{\psplit(N)} \! \mathrm{d}p \,  \frac{\Gsplit(p,N)}{H} \, n_\hard(N)
    \;.
\end{equation}
The integral is dominated by the UV modes with $p \sim \psplit$, 
\begin{equation}
    \label{eq:rho_soft_diffeq}
    \frac{d\rho_\soft}{d N} + 4 \rho_{\soft} \sim \psplit n_\hard \;. 
\end{equation}
Asymptotically $\psplit$ is dominated by the contribution from the soft particles themselves, and substituting $\psplit^\soft$ \eqref{eq:p_split} and $n_\hard$ \eqref{eq:rhoH_nH} yields
\begin{equation}
    \label{eq:rho_soft_diffeq_asymptotic}
    \frac{d\rho_\soft}{d N} + 4 \rho_{\soft} \sim A \mphi \rho_\soft^{3/4} e^{3 N/2} \;.
\end{equation}
Here we have used that $(n_\soft/g_*)^{1/3} \sim (\rho_\soft/g_*)^{1/4}$ for a thermal population, and we have defined a dimensionless constant 
\begin{equation}
    A \sim \alpha^4  \frac{g_\QCD}{g_*^{3/4}} \frac{\mphi}{\Hend} \Gtilde \;,
\end{equation}
which is much less than unity for the parameters of interest.  
The asymptotic solution of Eq.~\eqref{eq:rho_soft_diffeq_asymptotic} is 
\begin{equation}
    \label{eq:rhosoftasymptotic}
    \rho_\soft \sim \frac{A^4 \mphi^4}{10000} \, e^{6 N}\;. 
\end{equation}
Despite the cosmological redshifting, the thermal population's energy density increases with time. This growth continues until $\psplit$ reaches the hard particle energy $\mphi/2$ and $\rho_\soft \sim \rho_\hard$; this occurs 
\begin{equation}
    \label{eq:Ntherm}
    \Ntherm \sim \frac{2}{15} \ln\left[\frac{1}{\alpha^{16} \Gtilde^3} \frac{g_*^3}{g_\QCD^4}\left(\frac{\Hend}{\mphi}\right)^{5}\right]
\end{equation}
$e$-folds after the end of inflation. At this point, the inflaton's decay products (hard + soft) fully thermalize. The subsequent temperature is then set by conservation of energy,
\begin{align}
    \Tphi ( N > \Ntherm ) = \left( \frac{30 \rho_r}{\pi^2 g_*} \right)^{1/4} 
    \;,
\end{align}
and the plasma's maximum temperature is reached at the beginning of this fully thermal phase \cite{Harigaya:2013vwa}
\begin{align}\label{eq:Tmaxinflaton_explicit}
    \Tmax = \Tphi(\Ntherm) \sim \frac{\alpha^{4/5} \Gtilde^{2/5} m_\phi g_\QCD^{1/5}}{g_*^{2/5}}
    \;.
\end{align}
This is the maximum temperature of the inflaton decay products during reheating in the slow decay regime,  neglecting the presence and influence of the hot plasma produced by the Higgs. In Fig.~\ref{fig:temperature_evolution}, which we discuss further in the next subsection, it corresponds to the maximum of the purple line. 

\subsection{Effect of Higgs decay products}
\label{subsec:higgsoninf}

\begin{figure}[t]
\includegraphics[]{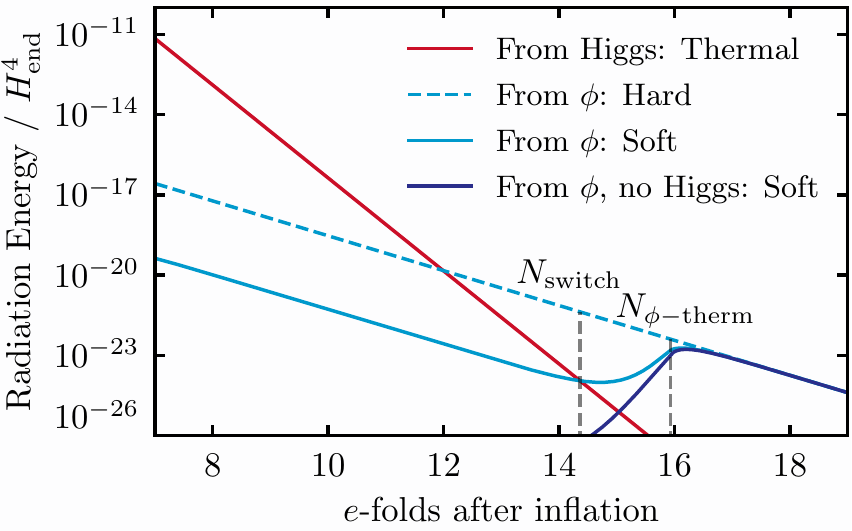}
\caption{The total energy of inflaton decay products after inflation (dashed blue, $\Gtilde=10^{-12}$) can only thermalize after hard primaries transfer their energy to a soft distribution by in-medium splitting. The Higgs ($\rhoparam = 10$) decays and produces an effectively thermal plasma early (red, $\Nthermhiggs = 6$), which adds additional scatterers into the medium and enables a soft population of inflaton decay products to form early (blue). By the time this soft population dominates over the Higgs ($N_{\rm switch}\sim14$), it will always have a lower energy density than it does once all the hard radiation thermalizes ($\Ntherm\sim16
$), which is the same as if there were no Higgs condensate at all (purple). See 
 \S\ref{subsec:higgsoninf} for further discussion.}
\label{fig:temperature_evolution}
\end{figure}

The Higgs condensate's decay products contribute to the medium in which the hard primaries split, amplifying the hard particles' splitting rate and changing their thermalization history. Previous studies of thermalization  during reheating, upon which \S \ref{subsec:cascade}-\ref{subsec:soft} are based, have neglected this Standard Model process.

As a simplifying assumption, we suppose that the decay of the Higgs condensate to an effectively thermal population of relativistic particles occurs abruptly at $\Nthermhiggs$. These decay products then provide scattering targets for the hard primaries, and the energy density of the soft radiation emitted by in-medium splitting $\rho_\soft$ evolves subject to an extension of Eq.~\eqref{eq:rho_soft_diffeq} where the splitting scale $\psplit$ is now
\begin{equation}
    \label{eq:p_split_with_h}
    \psplit = \psplit^\hard + \psplit^{\soft} + \psplithiggs\;.
\end{equation}
The first two terms appear in Eq.~\eqref{eq:p_split}, and the new contribution from the effectively thermal decay products of the Higgs is obtained by including their contribution to the elastic scattering rate \eqref{eq:Gamma_elastic} in the definition of the splitting scale \eqref{eq:Gamma_split}, yielding
\begin{equation}
    \label{eq:p_split_h}
    \frac{\psplithiggs}{\Hend} \sim \alpha^4 \frac{g_\QCD}{g_*^{3/4}} \rhoparam^{3/4} \;,
\end{equation}
which is time independent. 

Since the soft particles will themselves interact and thermalize with the Higgs' decay products, $\rho_\soft$ solved from Eq.~\eqref{eq:rho_soft_diffeq} should now be viewed as the energy transferred from the inflaton to the thermal plasma with total energy $\rho_\soft + \rho_\higgs$. We show the solution for $\rho_\soft$ for a typical parameter set in Fig.~\ref{fig:temperature_evolution}. We take an inflaton decay rate $\Gtilde = 10^{-12}$ and an inflaton mass $\mphi = \Hend$. We assume a Higgs energy $\rhoparam = 10$ and a Higgs decay time $\Nthermhiggs = 6$, which are typical values as seen in \eqref{eq:typical_rhoparam} and \eqref{eq:typical_Ndec}. We use a coupling strength $\alpha = 0.1$.

At early times, the contribution to the total plasma energy from the Higgs' decay products dominates over the contribution from all the inflation decay products. The splitting scale~\eqref{eq:p_split_with_h} is therefore dominated by $\psplit^{\rm higgs}$ and so the solution of Eq.~\eqref{eq:rho_soft_diffeq} is
\begin{equation}
\label{eq:rho_soft_with_higgs}
\rho_\soft \sim \min\left[\psplithiggs, \mphi \right] n_{\hard}\;,
\end{equation}
where the minimum in this expression limits splitting to occur below the energy of hard particles themselves.  Due to the $\alpha^4$ factor in \eqref{eq:p_split_h}, $\psplithiggs$ is smaller than $\mphi$ unless $\Hend \gg \mphi$ or $\rhoparam$ is much larger than unity.\footnote{While this can in principle be the case if $\lambda$ is very small (see Eq.~\eqref{eq:typical_rhoparam}), in the very small $\lambda$ regime the decay of the Higgs condensate by parametric resonance may be disrupted if the weak gauge bosons decay to fermions on the condensate oscillation timescale \cite{GarciaBellido:2008ab,Enqvist:2014tta}.}

Despite being enhanced by scatterings off the Higgs' decay products, $\rho_\soft$ 
is initially a subdominant component of the thermal plasma relative to the Higgs contribution $\rho_\higgs$. The solution \eqref{eq:rho_soft_with_higgs} is valid until $\rho_\soft$ 
overtakes $\rho_\higgs$ at time
\begin{equation}
    N_{\rm switch} \sim \frac{2}{5} \log\left( \frac{\rhoparam}{\Gtilde} \frac{\Hend}{\min\left[\psplithiggs, \mphi \right]} \frac{\Hend^2}{\mphi^2} \right)\;,
\end{equation}
which is the intersection time of the red and blue lines in Fig.~\ref{fig:temperature_evolution}. $N_{\rm switch}$ is an important time. Observables which depend on the temperature of the plasma during reheating, such as the dark matter relic abundance we will discuss in \S\ref{sec:implications}, inherit the fluctuations of the thermal plasma at the time they were produced. If they are produced before $N_{\rm switch}$, the Higgs contribution to the thermal plasma dominates and observables will inherit the Higgs' fluctuations. If they are produced after $N_{\rm switch}$, they inherit the inflaton's. We are therefore interested in the maximum temperature of the plasma before and after $N_{\rm switch}$.

Before $N_{\rm switch}$, the Higgs contribution dominates the thermal plasma and the maximum temperature is $\Tmaxhiggs$ \eqref{eq:Tmaxhiggs}.
At $N_{\rm switch}$, we write $\rhoparam$ in terms of $\psplithiggs$ to compare the energy of the thermal plasma to the maximum energy it has when it thermalizes without the Higgs $\rho_r(\Ntherm)$. We find
\begin{equation}
    \frac{\rho_{\rm switch}}{\rho_r(\Ntherm)} \sim \left(\frac{ \min\left[\psplithiggs, \mphi \right]^2}{\psplithiggs \mphi}\right)^{4/5} \lesssim 1\;,
\end{equation}
and thus regardless of whether $\psplithiggs$ is greater or less than $\mphi$, by the time the inflaton contribution to the thermal plasma dominates over the Higgs contribution, the plasma is inevitably at a lower temperature than its maximum without the Higgs, $\Tmax$.

After $N_{\rm switch}$, the energy $\rho_\soft$ transferred from the inflaton hard primaries dominates the energy density of the thermal plasma and therefore provides the dominant contribution to the splitting rate.  If $\psplithiggs$ was less than the hard particle energy $\mphi$, as we expect and as shown in Fig.~\ref{fig:temperature_evolution}, then $\psplit^{\soft}$ begins to grow as described in \S\ref{subsec:soft} and approaches the asymptotic solution \eqref{eq:rhosoftasymptotic}. Once $\rho_\soft$ reaches the hard particle energy $\rho_\hard$ the plasma thermalizes completely. At this point it reaches the maximum temperature $\Tmax$ computed without the effect of the Higgs in Eq.~\eqref{eq:Tmaxinflaton_explicit}.

We therefore see that while the Higgs can increase the hard primary splitting rate to enhance the inflaton contribution to the thermal plasma, $\Tmax$ still represents the maximum temperature of the plasma after $N_{\rm switch}$ as long as $\psplithiggs \lesssim \mphi$.

If, however, $\psplithiggs \gtrsim \mphi$, then the inflaton contribution to the thermal plasma only comes to dominate the Higgs one after the thermalization time $\Ntherm$ \eqref{eq:Ntherm} which defined the maximum temperature $\Tmax$ of the plasma neglecting the Higgs. Explicitly,
\begin{equation}
    N_{\rm switch}-\Ntherm \sim \frac{2}{15} \log\left(  \frac{\psplit^4}{\min\left[\psplithiggs, \mphi \right]^3} \frac{1}{\mphi} \right)\;.
\end{equation}
When $N_{\rm switch}>\Ntherm$, the maximum temperature of the plasma after the inflaton contribution dominates at $N_{\rm switch}$ is in fact determined by the energy at $N_{\rm switch}$ itself $\sim\rho_r(N_{\rm switch})$, lower than it was without the Higgs. We will not focus on this regime in the following. 

\section{Maximum Temperature during Reheating}
\label{sec:comparison}

We have seen that the thermal plasma during reheating receives contributions from the decay of the Higgs and inflaton condensates, and that it can be much hotter than the ultimate reheating temperature at the onset of radiation domination. The inflaton decay products  provide a maximum temperature \eqref{eq:Tmaxinflaton_explicit} 
\begin{align}\label{eq:Tphimax_over_Hend}
    \frac{\Tmax}{\Hend}& \sim 
    0.4\left( \frac{ m_\phi}{\Hend} \right)^{3/4} \tilde \Gamma^{1/4}
     e^{-3 \Ntherm/8}
    \nonumber\\
    & \sim
    10^{-3} \biggl(\frac{\mphi}{\Hend}\biggr) \biggl(\frac{\Gtilde}{10^{-4}}\biggr)^{2/5} \biggl(\frac{\alpha}{0.1}\biggr)^{4/5}\;,
\end{align}
which is controlled by the inflaton parameters $\mphi/\Hend$ and $\Gtilde$ and by the Standard Model parameters 
of which we have retained here only $\alpha$.

The decay of the Higgs condensate provides a maximum temperature \eqref{eq:typicalT}, 
\begin{equation}
\label{eq:Thmax_over_Hend}
    \frac{\Tmaxhiggs}{\Hend} 
    \sim 10^{-3}  \left(\frac{\rhoparam}{10}\right)^{1/4}
    e^{-(\Nthermhiggs-6)}
    \;,
\end{equation}
which is controlled by the Higgs density parameter $\rhoparam$ and by the Higgs effective thermalization time $\Nthermhiggs$. Both of these temperature scales can be much larger than the reheating temperature \eqref{eq:TRH} 
\begin{equation}
    \frac{\TRH}{\Hend} 
    \sim 10^{-7} \biggl( \frac{\Hend}{10^{10} \ \mathrm{GeV}} \biggr)^{1/2} \biggl( \frac{\mphi}{\Hend} \biggr)^{3/2} \biggl( \frac{\Gtilde}{10^{-4}} \biggr)^{1/2} 
\end{equation}
that characterizes the plasma at the onset of radiation domination.

\begin{figure}[t]
\includegraphics[]{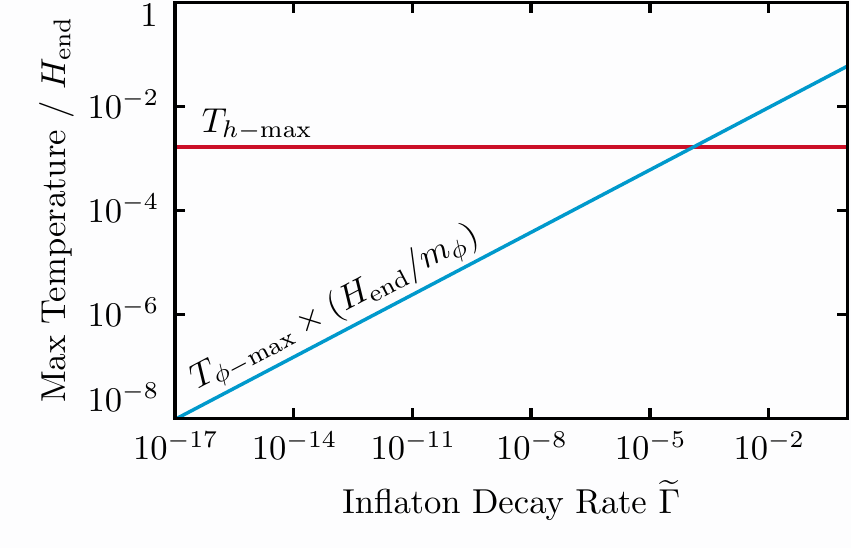}
\caption{The maximum temperature during reheating as a function of the inflaton decay rate $\Gtilde$. The Higgs decays to an effectively thermal population at a temperature $\Tmaxhiggs$ (red, \eqref{eq:typicalT}), which dominates over the contribution to the thermal plasma from the inflaton's thermal decay products $\Tmax$ (blue, \eqref{eq:Tmaxinflaton_explicit}) when $\Gtilde \lesssim 10^{-4}$ for $\Hend \sim \mphi$.
For further discussion see \S\ref{sec:comparison}.}
\label{fig:temperature}
\end{figure}

We show these temperatures in Fig.~\ref{fig:temperature} as a function of the inflaton decay rate $\Gtilde$ for the parameter values above. With these parameter choices the maximum temperature of our universe is provided not by the decay of the inflaton but by the decay of the Higgs condensate provided that the inflaton decay rate is sufficiently small,
\begin{equation}
\label{eq:gtildecondition}
\Gtilde \lesssim 10^{-4} \left(\frac{\Hend}{\mphi}\right)^{5/2} \;,
\end{equation} 
or equivalently when reheating takes a sufficiently long time
\begin{equation}
    \label{eq:NRH}
    N_{\rm RH} \gtrsim 32 - \frac{4}{3} \ln \left( \frac{\Hend}{10^{10} \GeV} \left(\frac{\mphi}{\Hend}\right)^{1/4} \right)\;.
\end{equation}

If the maximum temperature during reheating arises from the decay of the Higgs condensate, i.e.\ $\Tmaxhiggs > \Tmax$, then the plasma temperature will for some time track the Higgs condensate's spatial inhomogeneities.  Since these inhomogeneities result from the Higgs' stochastic fluctuations during inflation, the maximum temperature during reheating can therefore locally inherit large-scale spatial correlations.

In de~Sitter space, correlation functions must be invariant under the $\mathrm{SO}(4,1)$ isometry.  Consequently, the Higgs field's equal-time spatial correlation function $G(R)$ can be computed from the unequal-time temporal correlation function $G(\Delta N) \equiv \langle h(\vec{x}, N) \ h(\vec{x}, N+\Delta N) \rangle$. The leading order non-trivial behavior is \cite{Starobinsky:1994bd,Adshead:2020ijf}
\begin{equation}
    G(\Delta N) \propto  e^{-\vert\Delta N \vert/N_c}\;,
\end{equation}
with the asymptotic correlation time
\begin{equation}
    \label{eq:correlation_time}
    N_c \equiv \frac{H}{\Lambda_1} 
    \simeq \bigl( 110 \ e\textrm{-folds} \bigr) \left( \frac{\lambda}{10^{-2}} \right)^{-1/2}\;,
\end{equation}
determined by the smallest non-zero eigenvalue $\Lambda_1$ of the Fokker-Planck equation for the Higgs' one-point probability distribution function.\footnote{We use here the eigenvalue for a real field with a $Z_2$ symmetry, because only radial steps decohere the condensate. The variance of the $\mathrm{SU}(2)$-doublet Higgs is slightly larger relative to the radial steps than the variance of the $Z_2$ field, so the Higgs should be slightly harder to decohere and the correlation time \eqref{eq:correlation_time} is a mild underestimate.}

The equal-time spatial correlation function between points separated by a physical distance $R$ is then obtained by the mapping $\Delta N \rightarrow 2 \ln(R H)$. For regions probed by the CMB, which crossed the inflationary horizon some $60$ $e$-folds before the end of inflation, the physical distance is $R_{60} \sim H^{-1} e^{60}$, and we estimate the spatial correlation function to be 
\begin{equation}
    \label{eq:correlation}
    \frac{G(R_{60})}{G(0)} \sim e^{-2 \times 60/N_c} \sim 0.3
\end{equation}
for $\lambda = 10^{-2}$. 

These estimates imply that the Higgs condensate during reheating was only partially correlated on large length scales that correspond to our present Hubble patch.  If the Higgs condensate sets the maximum temperature of our universe during reheating, we expect order unity fluctuations in that temperature across the CMB scales.

Note that the evolution of the Hubble rate during inflation can change this expectation. As discussed in \S\ref{subsec:stochastic}, if the Hubble rate shrinks near the end of inflation then regions with stochastic Higgs fluctuations set earlier during inflation are released from Hubble drag and roll down the potential until their evolution is arrested once again. This causes regions with different values of the Higgs condensate to converge on the Hubble drag solution (see Fig.~\ref{fig:energy_density_during}). In the $\HdS/\Hend \gg 1$ limit, this attractor behavior enhances the Higgs' correlation on large scales relative to Eq.~\eqref{eq:correlation}, and can in principle reduce the inhomogeneity of the 
effects we will discuss in the next section.

\section{Implications for low-scale Reheating}
\label{sec:implications}

\subsection{Dark matter and unwanted relics}
\label{subsec:DM}

It may appear that by increasing the temperature of the primordial plasma the Higgs can source 
viable dark matter candidates when the inflaton alone cannot. This is not so. Any relic produced while the Higgs' decay products dominate the thermal plasma   inherits the Higgs' order unity correlations on large scales \eqref{eq:correlation}, which can be viewed as an extreme form of matter-radiation isocurvature fluctuations. 
One should therefore instead be concerned that scenarios with a low reheating temperature could generically suffer from the production of catastrophic relics sourced by the Higgs contribution to the thermal plasma. 
Fortunately, we shall see that this can only occur if $\Hend \gtrsim 10^{11} \GeV$, a region of parameter space which is likely excluded by the Higgs instability constraint $\Hend \lesssim \Hunstable \sim 10^{10} \GeV$.

Observations of the CMB constrain the amplitude of matter-radiation isocurvature perturbations $\mathcal{S}_{mr}$ to be $\lesssim 10^{-1} \mathcal{R}$ where $\mathcal{R}\simeq5\times10^{-5}$ is the amplitude of the adiabatic perturbations \cite{Akrami:2018odb}. Any stable, weakly-coupled relic $\chi$ that was produced by the thermal plasma in the epoch when it was dominated by the Higgs' decay products will carry $\mathcal{O}(1)$ energy inhomogeneities $\delta \rho_\chi \sim \rho_\chi$. If it is nonrelativistic at recombination it will behave as a dark matter component leading to an isocurvature amplitude $\mathcal{S}_{m r} \sim \delta \rho_\chi / \rho_\mathrm{tot}$ comparable to its relic abundance $\Omega_\chi \sim \rho_\chi/\rho_\mathrm{tot}$.
Consequently, the CMB isocurvature constraint implies an abundance constraint
\begin{equation}
\label{eq:isocurvature_bound}
\Omega_\chi h^2 \lesssim 10^{-6}\;.
\end{equation}

We now compute the relic abundance of particles produced during reheating by the thermal plasma. We separate the produced abundance into the contributions from the Higgs and inflaton decay products to see under what circumstances the bound \eqref{eq:isocurvature_bound} on the Higgs contribution is satisfied.

When the relic is produced during a matter-dominated reheating phase, its relic abundance can be related to its production time $N_*$ and its number density at production $n_\chi(N_*)$ through \cite{Giudice:2000ex}
\begin{equation}
\label{eq:abundance_general}
\Omega_\chi h^2 = \frac{\pi^2}{270} \frac{g_{*S,0} T_{\rm CMB}^3 m_\chi \TRH}{\Mpl^4 H_{100}^2} \left.\left(\frac{n_\chi}{H^2} \right)\right\vert_{N_*}\;,
\end{equation}
so long as after reheating the comoving entropy density is conserved. $m_\chi$ is the mass of the particle, $g_{*S,0} \simeq 3.91$ is the number of effective entropy degrees of freedom today, and $H_{100} = 100 \ \textrm{km/s/Mpc}$.

For simplicity we focus on thermal production mechanisms for $n_\chi$, though non-thermal production of relics during perturbative reheating can also generally be important due to the continual generation of hard primaries with energies $\sim\mphi$ \cite{Harigaya:2014waa,Garcia:2018wtq,Harigaya:2019tzu,Drees:2021lbm}. Since the Higgs provides additional scattering targets for those particles (see \S\ref{subsec:higgsoninf}), it may play a role in such processes as well.

The kinetic equation for $n_\chi$ is \cite{Giudice:2000ex}
\begin{equation}
\label{eq:kinetic}
\frac{d (e^{3 N} n_\chi)}{d N} = - \frac{\langle \sigma v \rangle}{H} e^{3 N } \left( n_\chi^2 - n_{\chi,{\rm eq}}^2 \right)\;,
\end{equation}
where the thermally-averaged annihilation cross-section $\langle \sigma v \rangle$ can be implicitly time dependent if it is temperature dependent, and $n_{\chi, {\rm eq}}$ is the equilibrium number density. We assume bosonic dark matter with one internal degree of freedom so $n_{\chi, {\rm eq}} = T^3 \zeta(3)/\pi^2$.

\subsubsection{Freeze out}
\label{subsubsec:FO}

Freeze out occurs {when $\langle \sigma v \rangle n_{\rm eq} \approx 3 H$, and if this occurs when $m_\chi \gg T$ then}  $n_\chi \gg n_{\chi,eq}$ and the kinetic equation \eqref{eq:kinetic} yields 
\begin{equation}
e^{3N} n_\chi (N) \simeq \left(\int_{N_{\rm FO}}^{N} d N \frac{\langle \sigma v \rangle}{H} e^{-3 N} \right)^{-1}\;, 
\end{equation}
where $N_{\rm FO}$ is the freeze-out time. So long as the cross section is not enhanced at low temperatures, the particle annihilation occurs mainly near $N_{\rm FO}$ and the number density at $N\gg N_{\rm FO}$ is simply 
\begin{equation}
n_\chi \simeq \frac{3}{2} \frac{\HFO}{\langle \sigma v \rangle_{\rm FO}} e^{-3(N-N_{\rm FO})} \;,
\end{equation}
where $\HFO$ is the Hubble rate when the freeze-out temperature $T_{\rm FO}$ is reached and $\langle \sigma v\rangle_{\rm FO}$ is the cross section at freeze out. The particular order unity numerical coefficient here assumes a temperature independent cross section.

First let us consider the well-known case (see, e.g., Refs.~\cite{Giudice:2000ex,Harigaya:2016vda}) in which the relic particle is produced by the inflaton decay products before radiation-domination is reached; this corresponds to $\TRH < T_{\rm FO} < \Tmax$.
Under these assumptions, the relic mass $m_\chi$, relic abundance $\Omega_\chi h^2$, freeze-out temperature $T_{\rm FO} \equiv m_\chi / \xFO$, reheating temperature $\TRH$, and scattering strength $\langle \sigma v \rangle$, are related through
\begin{equation}\label{eq:mchi_ov_TRH_inflaton_case}
\frac{m_\chi}{\TRH} \sim \frac{1}{\beta} \left(\frac{\TRH}{2 \TeV}\right)^2  \left(\frac{0.12}{\Omega_\chi h^2}\right) \left(\frac{\xFO}{20}\right)^{7/2}\;,
\end{equation}
where we have parameterized the cross section as
\begin{equation}
    \langle \sigma v \rangle_{\rm FO} = \beta \frac{4\pi}{m_\chi^2} {\sqrt{\frac{\xFO}{6}}}\;,
\end{equation}
which defines the dimensionless parameter $\beta$. The value $\xFO \sim 20$ is familiar from studies of weak-scale freeze out, and $\xFO$ is only logarithmically sensitive to the mass scale (see, e.g., \cite{Harigaya:2016vda}). 

The validity of the underlying assumptions requires that the cross section be sufficiently high that the particle reaches equilibrium before freezout but not so high as to violate unitary bounds. Unitarity requires $\beta \lesssim 1$, and thermal equilibrium requires $\langle \sigma v \rangle_{\rm FO} n_{\chi, \rm eq, FO} \gtrsim \HFO$, which translates to 
\begin{equation}
\label{eq:beta_lower} 
\beta \gtrsim \frac{\sqrt{6} \HFO \pi \xFO^{5/2}}{4 m_\chi \zeta(3)} \;.
\end{equation}
The upper and lower bounds on $\beta$ then yield, respectively, lower and upper bounds on the $m_\chi$ which can yield a given relic abundance at a given reheating temperature
\begin{align*}
\left(\frac{\TRH}{2 \TeV}\right)^{2} &\left(\frac{0.12}{\Omega_\chi h^2}\right) \left(\frac{\xFO}{20}\right)^{7/2}  \\ & \lesssim  \frac{m_\chi}{\TRH}  \lesssim \numberthis
 \\  10^4   &  \left(\frac{\TRH}{2 \TeV}\right)^{1/4}  \left(\frac{0.12}{\Omega_\chi h^2}\right)^{1/4}  \left(\frac{\xFO}{20}\right)^{5/4} \;,
\end{align*}
subject to the condition $\TRH < T_{\rm FO} < \Tmax$.  Notice that unitarity provides a \textit{lower bound} on the mass.  This is because $H \propto T^4$ before radiation domination; if instead freeze out occurs after radiation domination when $H \propto T^2$, then unitarity leads to the familiar upper bound on mass \cite{Griest:1989wd}.

For $\Omega_\chi h^2 \simeq 0.12$ such that $\chi$ can be all of the dark matter, 
the lower bound on the mass is below the upper bound for all reheating temperatures below $\sim 400 \TeV$. The crossing point determines the maximum mass particle which can be the dark matter, $m_\chi \sim 10^{10} \GeV$. On the other hand, in the limiting scenario of reheating right before BBN, $\TRH \sim 1 \MeV$, the maximum mass for a dark matter candidate is only $m_\chi \sim 300 \MeV$. 

Now if instead freeze out occurs at $ \Tmax < T_{\rm FO} < \Tmaxhiggs$, then the Higgs contribution to the plasma determines the relic abundance and we generically expect order unity inhomogeneities in its number density on large scales.

In this case the temperature is related to the Hubble rate as $H\propto T^{3/2}$ by Eq.~\eqref{eq:rhohiggs} and Eq.~\eqref{eq:H}, and so the abundance is enhanced for more massive particles. The analog of Eq.~\eqref{eq:mchi_ov_TRH_inflaton_case} is now
\begin{align*}
\frac{m_\chi}{\TRH} & \sim  \beta^{2/3} \left(\frac{10^{10} \GeV}{\Hend}\right)^{1/3}  \left(\frac{10}{\rhoparam} \right)^{1/4} \numberthis \\ & \quad \times \left(\frac{10^{7} \GeV}{\TRH}\right)^{5/3} \left(\frac{\Omega_\chi h^2}{10^{-6}}\right)^{2/3} \left(\frac{20}{\xFO}\right)^{2/3}\;, 
\end{align*}
where we have scaled $\Omega_\chi h^2$ by the isocurvature bound \eqref{eq:isocurvature_bound}. The lower and upper bounds on $\beta$ now provide, respectively, lower and upper bounds on the mass as\footnote{If the dark matter particles are relativistic at the time of freeze out, the resultant scaling would be the same as the lower limit on $m_\chi$ shown here.}
\begin{align*}
 & \left(\frac{ 10^{10} \GeV}{\Hend}\right) \left(\frac{10^{7} \GeV}{\TRH}\right)^2 \times \numberthis  \\ &\left(\frac{\Omega_\chi h^2}{10^{-6}}\right) \left(\frac{10}{\rhoparam}\right)^{3/4}  \lesssim \frac{m_\chi}{\TRH} \lesssim  \left(\frac{\Omega_\chi h^2}{10^{-6}}\right)^{2/3} \left(\frac{10}{\rhoparam}\right)^{1/4}  \\   &\qquad  \qquad \times  \left(\frac{10^{10} \GeV}{\Hend}\right)^{1/3} \left(\frac{10^{7} \GeV}{\TRH}\right)^{5/3}  \left(\frac{20}{\xFO}\right)^{2/3} \;, 
\end{align*}
subject to the condition  $\Tmax < T_{\rm FO} < \Tmaxhiggs$. The allowed region increases  for increasing $\Hend$ and $\TRH$. Replacing $\TRH$ with $\Gtilde$ and $\mphi$, then $\Gtilde < 10^{-4}$  \eqref{eq:gtildecondition} sets a lower bound on $\Hend$ for which a solution for a given 
relic abundance exists
\begin{align*}
\label{eq:danger_zone_FO}
\Hend \gtrsim&\  10^{11} \GeV  \left(\frac{\Omega_\chi h^2}{10^{-6}}\right)^{2/7} \left(\frac{10^{-4}}{\Gtilde}\right)^{1/7} \\ &\times \left(\frac{10}{\rhoparam}\right)^{3/7} \left(\frac{\xFO}{20}\right)^{4/7} \left(\frac{\Hend}{\mphi}\right)^{3/7}\;. \numberthis
\end{align*}

This limit sets the Hubble rate at the end of inflation above which the presence of the Higgs can be dangerous. If $\Hend$ is above $10^{11} \GeV$ and $\Gtilde$ is suitably small but not too small (though the dependence is very weak), then the Higgs condensate can source relics by thermal freeze out in sufficient abundance to interfere with the CMB.
Conversely if $\Hend \ll 10^{11} \GeV$ then no such relics can form.  

Remarkably, this safe region is similar to the Higgs instability bound, which restricts the inflationary Hubble rate to be below $\Hunstable \sim 10^{10} \GeV$. Thus if the Hubble rate during inflation is low enough that the Higgs was not sent to its instability, which we implicitly assume in our calculations, then by this criteria
the Higgs is also not able to source dangerous isocurvature perturbations by thermal freeze out.  The exact bound $\Hunstable$ is on the other hand currently subject 
to experimental uncertainties in the top mass, as well as the stochastic history of the Higgs during inflation, 
and our results therefore provide an independent mechanism by which the Higgs may limit the inflationary energy scale.

\subsubsection{Freeze in}

The lower bound on the cross section \eqref{eq:beta_lower} is the condition that the relic was once in equilibrium with the thermal bath. When relic particles are produced from the bath at a low enough rate that the process was never in equilibrium, thermal freeze in occurs instead~\cite{Kusenko:2006rh,deGouvea:2006wd,Gopalakrishna:2006kr,Petraki:2007gq,Page:2007sh,Kusenko:2009up,Hall:2009bx,Kolb:2017jvz}.

With $n_\chi \ll n_{\chi,{\rm eq}}$, the kinetic equation \eqref{eq:kinetic} yields
\begin{equation}
e^{3N} n_\chi (N) \sim \frac{g_\chi^2 \zeta(3)^2}{\pi^4} \int_0^N dN \frac{e^{3 N} T^6 \langle \sigma v \rangle}{H}\;,
\end{equation}
where we assume that $m_\chi \ll T(N)$ while particle production occurs. Freeze-in production is inefficient in the complementary regime, $m_\chi \gtrsim T(N)$, where the rate is Boltzmann suppressed.
The integral may be dominated by early times or late times, depending on how $T$, $H$, and $\langle \sigma v \rangle$ vary in time.

With $H\propto T^{3/2}$ the Higgs contribution comes from production at $\Tmaxhiggs$ for all $\langle \sigma v \rangle$ which do not decrease with increasing temperature. This UV-dominated freeze in yields a number density at production \cite{Elahi:2014fsa} 
\begin{equation}
n_\chi (\Nthermhiggs) \simeq  \frac{g_\chi^2 \zeta(3)^2}{\pi^4}  \frac{\Tmaxhiggs^6 \langle \sigma v \rangle_{h\text{-max}}}{H(\Nthermhiggs)}\;,
\end{equation}
where $\langle \sigma v \rangle_{h\text{-max}}$ is the production rate at $\Tmaxhiggs$ and the specific order unity coefficient here assumes a constant $\langle \sigma v \rangle$. 

With $H\propto T^4$ the inflaton contribution, on the other hand, comes from production at the end of reheating for constant $\langle \sigma v \rangle$. This remains true for UV-enhanced cross sections $\langle \sigma v \rangle\propto T^n$ for all $n < 6$ \cite{Garcia:2017tuj}. Only for $n>6$ is the dominant contribution from $\Tmax$. Regardless, the relative relic abundance produced by the Higgs and inflaton depends on the relic mass and the relative values of $\langle \sigma v \rangle $ at the respective production times. We focus here on the regime where $\langle \sigma v \rangle $ is sufficiently larger at $\Tmaxhiggs$ that the Higgs decay products provide the dominant contribution to the abundance.

That abundance is then bounded from above by the condition that it was not in thermal equilibrium,
\begin{equation}
\langle \sigma v \rangle_{h\text{-max}} n_{eq, h\text{-max}} < H_{h\text{-max}}\;, 
\end{equation}
which is stronger than the relativistic version of the unitarity bound on the cross-section. As in the freeze-out case, the bound can be translated into a bound on $\Hend$ given a relic density 
\begin{align*}
\label{eq:danger_zone_FI}
\Hend &\gtrsim\numberthis  10^{11} \GeV \left(\frac{\Omega_\chi h^2}{10^{-6}}\right)^{2/7} \left(\frac{10^{-4}}{\Gtilde}\right)^{1/7} \\ &\left(\frac{\Tmaxhiggs}{m_\chi}\right)^{2/7} \left(\frac{10}{\rhoparam}\right)^{3/7} \left(\frac{\Hend}{\mphi}\right)^{3/7}\;,
\end{align*}
where we have scaled $m_\chi$ by $\Tmaxhiggs$ but note it should satisfy $m_\chi \ll \Tmaxhiggs$ for our approximations to be consistent.
The danger zone for potentially producing isocurvature perturbations that are too large is quite similar to the freeze-out one \eqref{eq:danger_zone_FO}:
$\Hend \gtrsim 10^{11} \GeV$.  

\subsection{Symmetry Restoration}

The maximum temperature of the thermal plasma during reheating has implications for the restorations of symmetries by finite temperature effects. The maximum temperature of the thermal plasma neglecting the presence of the Higgs condensate \eqref{eq:Tphimax_over_Hend} can be written in terms of the reheating temperature as \cite{Harigaya:2013vwa,Mukaida:2015ria}
\begin{equation}
\label{eq:Tmaxinflaton_RH}
\Tmax \sim 100 \GeV \left(\frac{10^{10} \GeV}{\mphi}\right)^{1/5} \left(\frac{\TRH}{1 \MeV}\right)^{4/5}\;.
\end{equation}
Note that for a fixed reheating temperature it is inversely dependent on $\mphi$. The electroweak symmetry is restored if the temperature reaches values $\gtrsim 100 \GeV$, and we therefore see that, neglecting the Higgs condensate, for low reheating temperatures the electroweak symmetry is not necessarily restored by the thermal plasma if $\mphi \gtrsim 10^{10} \GeV$. This point was discussed in Ref. \cite{Mukaida:2015ria}.

The decay of the Higgs condensate, however, provides a plasma temperature 
\eqref{eq:Thmax_over_Hend} 
\begin{equation}
\label{eq:typicalT_explicit}
    \Tmaxhiggs 
    \sim 10^{7} \left(\frac{\Hend}{10^{10} \GeV }\right)  \left(\frac{\rhoparam}{10}\right)^{1/4}
    e^{-(\Nthermhiggs-6)}
    \;,
\end{equation}
independent of the reheating temperature. Comparing the maximum temperature from the Higgs decay to the maximum temperature from the inflaton decay, we see that if $\mphi\sim\Hend$ then the decay of the Higgs is complementary to the decay of the inflaton: whenever the electroweak symmetry is not restored by the inflaton it is restored by the Higgs.

Thanks to the decay of the Higgs condensate, the electroweak symmetry is therefore in fact guaranteed to be restored in the early universe when reheating is perturbative, except when the Hubble rate at the end of inflation is low,
\begin{equation}
\Hend < 10^5 \GeV \left(\frac{0.01}{\lambda}\right)^{1/2} \left(\frac{10}{\rhoparam}\right)^{3/4},
\end{equation}
and much smaller than the inflaton mass
\begin{equation}
\frac{\mphi}{\Hend} > 10^5 \left(\frac{\lambda}{0.01}\right)^{1/2} \left(\frac{\rhoparam}{10}\right)^{3/4} \left(\frac{\TRH}{1 \MeV}\right)^4.
\end{equation}

The Higgs condensate may similarly play a role in the restoration of symmetries predicted by theories of new physics.  For instance the hypothetical Peccei-Quinn axion arises as the pseudo-Nambu-Goldstone boson associated with the spontaneous breaking of a global $\mathrm{U}(1)_\mathrm{PQ}$ symmetry \cite{Peccei:1977hh,Peccei:1977ur,Weinberg:1977ma,Wilczek:1977pj}.  
If the primordial plasma temperature initially exceeds the symmetry breaking scale, then as the universe cools the symmetry is broken in a cosmological phase transition which can lead to the formation of topological defects such as axion strings and domain walls. These can leave distinctive imprints on cosmological observables \cite{Marsh:2015xka}. The symmetry breaking scale is model-dependent, and some of the most compelling scenarios have $\sim 10^{10} - 10^{12} \ \mathrm{GeV}$ 
\cite{DiLuzio:2020wdo}, but even if inflation occurs at a high energy scale the inflaton decay products may not be hot enough to restore such symmetries if the inflaton decay is slow \cite{Mukaida:2015ria}. The decay of the Higgs condensate, on the other hand, leads to a high temperature plasma irrespective of the reheating temperature and may facilitate such symmetry restorations and the accompanying phase transitions and topological defect formation. 
However, unless the Peccei-Quinn scale is sufficiently low, i.e. $\lesssim \Tmaxhiggs$ from Eq.~\eqref{eq:typicalT_explicit}, even the Higgs condensate decay will be insufficient to restore the symmetry. 

Note that the large scale inhomogeneity of $\Tmaxhiggs$ does not directly impact relic defects from symmetry breaking except in rare regions where $\Tmaxhiggs$ fluctuates to such low values that the given symmetry is not restored. 

\section{Conclusion}
\label{sec:conclusion}

We have studied the role of the Standard Model Higgs field during the epoch of reheating after inflation. The Higgs forms a condensate during inflation, and we have focused on its behavior in the scenario where the inflaton decays slowly, having a small perturbative decay rate $\Gphi \ll m_\phi^3 / \Mpl^2$. We have calculated the maximum temperature achieved by the primordial plasma due to energy transfer from the Higgs and inflaton condensates, and the time evolution of this temperature between the end of inflation and the start of radiation domination.  

To do so, we built upon well-established results in the literature and applied them to the Higgs, including the formation of a Higgs condensate during inflation from its quantum fluctuations as a light spectator field, the dynamics of the Higgs condensate after inflation as it oscillates on its quartic potential, and the parametric resonance in the electroweak gauge fields that induces an energy transfer from the Higgs condensate into an effectively thermal plasma which can be understood qualitatively analytically and quantitatively by lattice simulations.

We contrasted the resonant decay of the Higgs condensate with the inflaton's perturbative decay into rare hard particles which
thermalize by an in-medium splitting
suppressed by the LPM effect, 
in doing so drawing from the extensive
literature on thermalization in non-abelian plasmas in the effective kinetic theory.

From this synthesis of constituent ideas, we are able to develop a comprehensive understanding of the Higgs condensate's role in reheating. The central conclusions of our work are: 
\begin{enumerate}
    \item Regardless of the inflationary history, the Higgs condensate has a typical energy density after inflation which lies within a narrow window between $\sim \lambda^{-2/3} \Hend^4 e^{-4N}$ and $\sim \lambda^{-1} \Hend^4 e^{-4 N}$ controlled by the Hubble rate at the end of inflation.
    
    \item  The Higgs condensate is important to the thermal history of the universe when the inflaton decay rate is sufficiently small \eqref{eq:gtildecondition},
    
    \hfil$
        \Gtilde \equiv \Gphi \Mpl^2 / m_\phi^3 \lesssim 10^{-4} \left(\dfrac{\Hend}{\mphi}\right)^{5/2},
    $\hfill

    or equivalently when reheating takes a sufficiently long time, see Eq.~\eqref{eq:NRH}.

    \item  The maximum temperature of the primordial plasma is then obtained at the time of the Higgs condensate's fragmentation and decay, typically a few $e$-folds after the end of inflation, and for typical parameters in Eq.~\eqref{eq:typicalT_explicit} is 
    
    \hfil$
        \Tmaxhiggs \sim 10^7 \GeV \left(\dfrac{\Hend}{10^{10} \GeV}\right) \;.
    $\hfill
    
    In the absence of a Higgs condensate, by contrast, the maximum temperature could be much lower, see Eq.~\eqref{eq:Tmaxinflaton_RH}.
    
    \item As shown in Fig.~\ref{fig:temperature_evolution}, the presence of the Higgs' decay products changes the thermalization history of the hard primaries produced as the inflaton decays by increasing the scattering targets in the plasma. This enhances the energy transfer from the hard primaries to the thermal sector, but once the inflaton contribution dominates the thermal energy of the universe the subsequent maximum temperature is set by the Higgs-less result above.

    \item The maximum temperature of our universe will inherit the Higgs' large scale stochastic fluctuations, which are uncorrelated with the curvature fluctuations in our universe. If relics are produced in substantial abundance from the plasma while it is dominated by the Higgs' decay products, they will therefore lead to unacceptable isocurvature fluctuations in the CMB. We show that for production by either thermal freeze out \eqref{eq:danger_zone_FO}  or freeze in \eqref{eq:danger_zone_FI} this cannot occur so long as the inflationary scale is below 
    
    \hfil $\Hend \lesssim 10^{11} \GeV.$\hfill
    
    \item Even for low reheating temperatures, the electroweak symmetry in our universe is  restored by the decay of the Higgs condensate after inflation if $\Hend \gtrsim  10^5 \GeV$, which complements symmetry restoration from the inflaton decay products.  Peccei-Quinn axion symmetry may also be restored if its symmetry breaking scale is below $\Tmaxhiggs$.

\end{enumerate}

\vspace*{1cm}

\acknowledgments

We thank Peter Adshead, Keisuke Harigaya, Kyohei Mukaida, David Weir, and Masaki Yamada for fruitful discussions.  We are especially grateful to Mustafa Amin and Kaloian Lozanov for illuminating discussions of preheating and lattice simulation.  

S.P was supported by the World Premier International Research Center Initiative (WPI), MEXT, Japan. W.H. was supported by U.S. Dept. of Energy contract DE-FG02-13ER41958 and the Simons Foundation.
D.Z. was supported by the National Science Foundation Graduate Research Fellowship Program under Grant No. DGE 1746045. 
\vfill

\bibliographystyle{apsrev4-1}
\bibliography{references.bib}

\end{document}